\documentclass[conference]{IEEEtran}
\IEEEoverridecommandlockouts

\usepackage{xcolor}
\usepackage{xspace}
\usepackage{graphicx}
\usepackage{subcaption}
\usepackage{booktabs}
\usepackage{multirow}
\usepackage{amsmath}
\usepackage{amssymb}
\usepackage{amsthm}
\usepackage{algorithm}
\usepackage{algorithmicx}
\usepackage[noend]{algpseudocode}
\usepackage{enumitem}
\usepackage[hidelinks]{hyperref}
\usepackage{balance}

\newcommand{\REMOP}{\textnormal{\textsc{REMOP}}\xspace}
\newcommand{\REMON}{\textnormal{\textsc{REMON}}\xspace}
\newcommand{\Description}[1]{}

\newtheorem{theorem}{Theorem}
\newtheorem{definition}[theorem]{Definition}
\newtheorem{property}[theorem]{Property}

\title{REMOP: REmote-Memory-aware \\ OPerator Optimization}

\author{
\IEEEauthorblockN{
Shiquan Zhang\IEEEauthorrefmark{1},
Yunhao Mao\IEEEauthorrefmark{1},
Yuqiu Zhang\IEEEauthorrefmark{2},
Gengrui Zhang\IEEEauthorrefmark{3},
Jeyhun Karimov\IEEEauthorrefmark{4},
Hans-Arno Jacobsen\IEEEauthorrefmark{1}}
\IEEEauthorblockA{%
\IEEEauthorrefmark{1}University of Toronto \quad
\IEEEauthorrefmark{2}Confluent Inc. \quad
\IEEEauthorrefmark{3}Concordia University \quad
\IEEEauthorrefmark{4}Microsoft Corporation\\
\{shiquan.zhang, yunhao.mao\}@mail.utoronto.ca, jacobsen@eecg.toronto.edu,\\
qzhang@confluent.io, \quad
gengrui.zhang@concordia.ca, \quad jkarimov@microsoft.com}
\thanks{This work has been submitted to the IEEE for possible publication. Copyright may be transferred without notice, after which this version may no longer be accessible.}
}

\begin{document}
\maketitle

\begin{abstract}
Remote and disaggregated memory tiers expand the effective memory capacity of analytical database engines, but they also reshape the cost structure of out-of-memory query processing.
When an operator spills beyond local DRAM, moving pages to remote memory incurs both data-transfer time and a fixed round-trip latency per transfer.
Classical operator analyses and buffer-allocation heuristics primarily target disk spilling by minimizing total I/O volume.
Under remote memory, these strategies can be suboptimal because they may trigger excessive transfer rounds.
We present \REMOP, a remote-memory-aware operator optimization framework that uses transfer-round-aware intra-operator memory policies to improve out-of-memory execution under tight memory budgets.
\REMOP introduces the number of transfer rounds into the latency cost model and derives operator-specific buffer-partitioning strategies, instantiating the approach for blocked nested-loop join, external merge sort, and external hash join in DuckDB.
Our evaluation on a two-node compute--memory testbed shows that \REMOP reduces transfer rounds by up to $97\%$ and operator runtime by up to $48\%$ on spill-heavy microbenchmarks, and lowers the average runtime of spilling TPC-H and TPC-DS queries by $22.7\%$ and $26.4\%$ end-to-end.
\end{abstract}



\section{Introduction}
\label{sec:intro}


Modern analytical workloads often approach the memory limits of database engines~\cite{Leis2018LeanStore, Neumann2020Umbra, DeBrabant2013AntiCaching, Otaki2025PagedMemoryManagement}.
Concurrency and pipeline parallelism further constrain per-operator budgets, causing operators to materialize working sets that exceed local DRAM~\cite{MehtaDeWitt1993, NagDeWitt1998, DavisonGraefe1994}.
The engine must then swap pages between DRAM and a slower tier. In such out-of-memory executions, the time spent swapping pages can substantially affect end-to-end latency~\cite{Neumann2020Umbra, DeBrabant2013AntiCaching, Otaki2025PagedMemoryManagement, Graefe1993Survey, Shapiro1986Joins}, making performance sensitive to how operators batch and schedule swaps.

Decades of out-of-memory query-processing research have developed operator analyses and memory-allocation heuristics for disk spilling~\cite{Graefe1993Survey, Shapiro1986Joins, Graefe2006Sorting, Knuth1998TAOCP3}.
These techniques (e.g., multi-pass external merge sort) largely target the bandwidth-dominated regime by reducing \emph{total I/O volume} and promoting sequential access, which is a useful proxy when per-I/O overhead is small relative to transfer time.

Recently, memory disaggregation has emerged as a practical way to extend the effective memory capacity of database engines via RDMA, CXL, and network-attached memory pools~\cite{Ruan2020AIFM, Shan2018LegoOS, Pond2023, Ahn2024CXLHANA}.
In this setting, spilled pages are placed in remote memory rather than on the local disk.
However, remote memory changes the cost structure of out-of-memory query processing: beyond data-transfer time, each swap event incurs a fixed end-to-end round-trip overhead due to network traversal, protocol processing, and remote-side memory access.

Table~\ref{tab:tier_numbers} summarizes representative bandwidth and latency across common swapping media~\cite{Bai2025MuScope, DeanLatencyNumbers, IntelX550ProductBrief, MicronDDR4UDIMM, MicrowayNetworkFabrics, SeagateNytro1370Manual}.
In particular, remote memory can provide higher bandwidth (BW) than SSDs, yet TCP/IP access adds hundreds of microseconds of round-trip time (RTT), often exceeding SSD access latency.
This fixed cost makes the \emph{number of swap events}, or transfer rounds, a first-order performance factor: moving the same bytes in more rounds can be slower because every round pays the RTT.
Thus, conventional disk-oriented allocations that minimize I/O volume can be suboptimal under remote memory when they induce excessive rounds.

To address this issue, we present \REMOP, a remote-memory-aware operator optimization framework that tunes intra-operator buffer allocation under tight memory budgets.
Its latency model balances data volume against transfer rounds, allowing modestly more data movement when this substantially reduces round trips.
We instantiate \REMOP for blocked nested-loop join, external merge sort, and external hash join, and implement it in DuckDB through a lightweight memory policy module and optimized operator variants.
Our evaluation shows that \REMOP can drastically reduce transfer rounds, yielding substantial operator-level speedups and consistent end-to-end improvements under tight memory limits.
Overall, we make three main contributions:

{\renewcommand{\labelenumi}{(\roman{enumi})}%
\begin{enumerate}[wide=0pt,labelsep=0.35em]
    \item We introduce a latency cost model for out-of-memory execution that accounts for both data volume and transfer rounds, and a lightweight operator abstraction for tuning intra-operator buffer partitions under remote memory (\S\ref{sec:fm}).
    \item As case studies, we instantiate \REMOP on blocked nested-loop join (BNLJ), external merge sort (EMS), and external hash join (EHJ), and derive network-adaptive buffer-allocation strategies. For BNLJ and EMS, these reduce transfer rounds by a factor of $\Theta(M)$ (on the order of the memory budget) and by up to 25\% over conventional allocations, respectively (\S\ref{sec:algo}).
    \item We implement \REMOP in DuckDB over TCP-based \REMON\footnote{\REMON~\cite{remon_icde2026} is our in-house remote-memory prototype.} and RDMA-based Infiniswap, and evaluate it with microbenchmarks and analytical workloads. Compared with vanilla DuckDB, \REMOP reduces transfer rounds by up to $97\%$ and operator runtime by up to $48\%$ in single-operator queries, lowers the average runtime of spilling TPC-H and TPC-DS queries by $22.7\%$ and $26.4\%$, and achieves gains up to $51\%$ under higher remote-memory latency (\S\ref{sec:impl}, \S\ref{sec:eval}).
\end{enumerate}}

The rest of the paper is organized as follows.
Section~\ref{sec:fm} presents the cost model and framework, and Section~\ref{sec:algo} derives buffer allocations for BNLJ, EMS, and EHJ.
Sections~\ref{sec:impl} and~\ref{sec:eval} describe the implementation and evaluation, followed by related work and conclusions in Sections~\ref{sec:rel_work} and~\ref{sec:conclu}.

\begin{table}[t]
  \centering
  \caption{Representative order-of-magnitude bandwidth and latency numbers across storage and memory tiers.}
  \label{tab:tier_numbers}
  \begin{tabular}{lcc}
    \toprule
    Medium & Bandwidth & Latency/RTT \\
    \midrule
    DRAM (DDR4) & $\sim$25.6\,GB/s & $\sim$100\,ns \\
    SSD (NAND) & $\sim$0.53\,GB/s & $\sim$100\,$\mu$s \\
    TCP/IP (intra-DC) & $\sim$1.25\,GB/s & $\sim$500\,$\mu$s \\
    RDMA (InfiniBand) & $\sim$6.8\,GB/s & $\sim$1\,$\mu$s \\
    \bottomrule
  \end{tabular}
\end{table}


\section{Cost Model and Optimization Framework}
\label{sec:fm}


Remote memory expands the effective capacity of database engines, but changes the cost of out-of-memory processing once queries spill beyond local DRAM.
Unlike bandwidth-bound disk I/O, remote memory offers high bandwidth but incurs a non-negligible RTT per request.
Operator performance therefore depends on both the \emph{total data volume} and the \emph{number of transfer rounds}.
This section formalizes this cost model (\S\ref{subsec:lat_model}), introduces \REMOP's operator optimization framework (\S\ref{subsec:mem_model}), and illustrates its tuning opportunities through concrete examples (\S\ref{subsec:example}).

\subsection{Cost Model with Remote Memory}
\label{subsec:lat_model}

Classical cost models for out-of-memory operators count pages transferred between memory and secondary storage~\cite{Knuth1998TAOCP3, Graefe2006Sorting, Shapiro1986Joins}.
This captures disk-based systems, where sequential bandwidth is the bottleneck and I/O cost is roughly proportional to data volume, but not the different cost structure of remote memory.

Remote memory over TCP/IP or RDMA can match or exceed SSD bandwidth, but each access incurs a fixed RTT from network traversal, protocol processing, and remote-side access~\cite{Ruan2020AIFM, Shan2018LegoOS, MicrowayNetworkFabrics}.
As Table~\ref{tab:tier_numbers} shows, TCP/IP provides gigabyte-scale bandwidth, while end-to-end RTT ranges from hundreds of microseconds within a datacenter to milliseconds across more remote deployments.

Figure~\ref{fig:lat_model} contrasts the latency models of disk-based and remote-memory swapping.
Despite its higher bandwidth, remote memory can have a longer RTT, making latency sensitive to the \emph{number of swap events}, or transfer rounds, because each round pays a fixed cost regardless of payload size.
Specifically, the overall swapping latency can be expressed as:

\begin{figure}[t]
  \centering
  \includegraphics[width=0.6\linewidth]{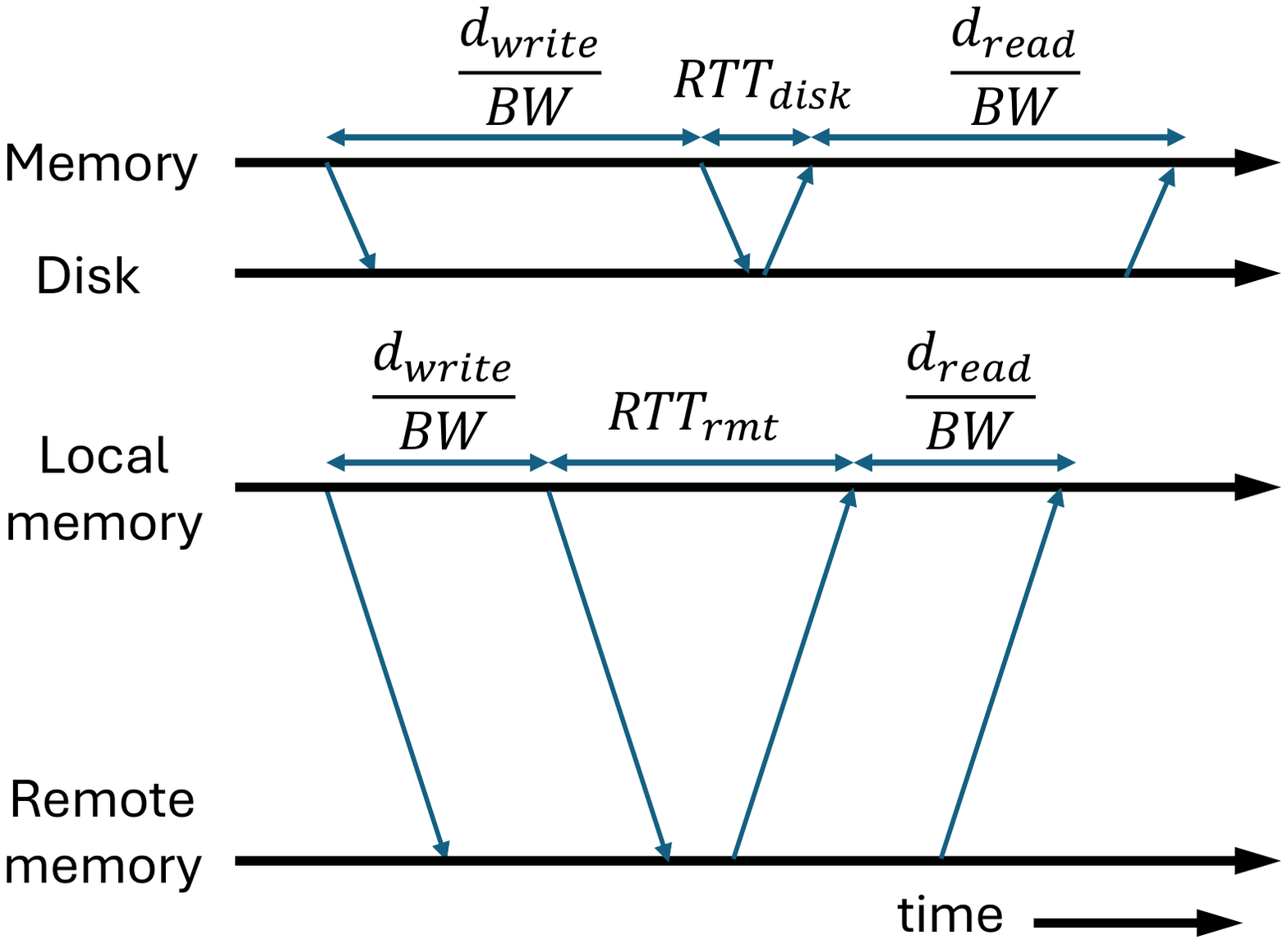}
  \caption{Latency model for disk vs. remote memory swaps.}
  \Description{A comparison of disk-based and remote-memory swapping latency, showing that remote memory adds a fixed per-transfer round-trip overhead.}
  \label{fig:lat_model}
\end{figure}

\begin{equation}
  \text{Latency} = \sum_{i=1}^{C} (\frac{d_i}{BW} + RTT) = \frac{D}{BW} + C \cdot RTT
  \label{eq:latency}
\end{equation}
where $d_i$ is the data volume of each swap, $D$ is the total data volume, $C$ is the number of transfer rounds, and $BW$ and $RTT$ capture the network conditions.

In classical disk spilling, operators amortize fixed overhead through large sequential transfers, so runtime often follows the volume term $D/BW$~\cite{Graefe1993Survey, Graefe2006Sorting, Shapiro1986Joins}.
Under remote memory, however, the round term $C\cdot RTT$ can also become substantial~\cite{Ruan2020AIFM, Infiniswap2017, Shan2018LegoOS}.
For example, swapping $D{=}10$\,GB in $C{=}20{,}000$ rounds costs about $D/BW{\approx}19$\,s for data transfer and $C\cdot RTT{\approx}2$\,s for round overhead on an SSD.
In contrast, over TCP/IP, the corresponding costs are about $8$\,s and $10$\,s, so the round term can dominate.
This creates a trade-off: reducing $C$ may lower total swapping time even if it increases $D$.

\subsection{Operator Optimization Framework}
\label{subsec:mem_model}

\REMOP is a remote-memory-aware operator optimization framework for the regime where each batched transfer incurs non-negligible RTT.
We now define an operator abstraction that captures batched swapping behavior and exposes the buffer-allocation knobs that determine $C$ and $D$.
Table~\ref{tab:notation} summarizes the notation and the scope of each parameter.

Consider an operator with an $M$-page memory budget that processes data exceeding $M$.
It partitions the budget into an input region $R_{in}=r_{in}M$ and an output region $R_{out}=r_{out}M$, with $r_{in}+r_{out}\approx1$ as data blocks dominate runtime bookkeeping.
The operator repeatedly loads up to $R_{in}$ pages and flushes up to $R_{out}$ pages when the output buffer fills.
Each such batched swap-in or flush-out is a \emph{transfer round}.
This abstraction broadly covers operators whose memory is divided into a few buffer regions and whose out-of-memory execution performs repeated batched reads and writes.

\begin{definition}[Data Transfer Cost]
For an operator that reads $D_{read}$ total pages and writes $D_{write}$ total pages, the \emph{total data transferred} (in pages) is $D = D_{read} + D_{write}$.
\end{definition}

\begin{definition}[Transfer Round Cost]
For an operator that performs $C_{read}$ read rounds and $C_{write}$ write rounds, the total number of \emph{transfer rounds} is $C = C_{read} + C_{write}$.
\end{definition}

The \emph{data transfer cost} $D$ corresponds to the classical I/O-volume measure and is mainly determined by the algorithm and data size.
The \emph{transfer round cost} $C$, in contrast, depends more on how $M$ is partitioned across $R_{in}$/$R_{out}$ and how $R_{in}$ is split among inputs.
From Eq.~\eqref{eq:latency}, we now define the \emph{total latency cost}:
\begin{definition}[Total Latency Cost]
For an operator that transfers $D$ pages in $C$ rounds, the \emph{total latency cost} is
\begin{equation}
  \label{eq:lat_cost}
  L \;=\; D + \tau\, C, \qquad \tau \propto BW \cdot RTT,
\end{equation}
\end{definition}
where $\tau$ is the network parameter proportional to $BW \cdot RTT$, setting the relative weight of the two terms.
When $\tau \to 0$ (e.g., bandwidth-bound local disk), $L$ reduces to the classical I/O-volume objective $\min D$; when $\tau$ is large (e.g., RTT-bound remote memory), the round term dominates and the objective approaches $\min C$.

Given a fixed budget $M$, \REMOP minimizes $L$ by tuning the input/output split $R_{in}$:$R_{out}$ and, when applicable, the split among multiple inputs or runs.
Under memory pressure, operators commonly follow a batched \emph{read\textendash{}compute\textendash{}write} loop.
\REMOP sizes the buffer partitions to increase the work per swap-in or flush-out, reducing $C$, and prefetches the next batch to overlap transfer with computation.
Our implementation realizes both mechanisms through operator-specific buffer sizing and pin/unpin behavior controlled by a lightweight policy module (\S\ref{sec:impl}).
These intra-operator optimizations complement plan-level memory budgeting~\cite{NagDeWitt1998} and buffer-manager replacement~\cite{Leis2018LeanStore, Neumann2020Umbra}.

\begin{table}[t]
  \centering
  \caption{Notation summary; sizes are in pages unless noted.}
  \label{tab:notation}
  \setlength{\tabcolsep}{4pt}
  \begin{tabular}{clc}
    \toprule
    Symbol & Description & Scope \\
    \midrule
    $M$       & Total operator memory budget & \REMOP \\
    $N$       & Total data size & \REMOP \\
    $D$       & Total data transferred & \REMOP \\
    $C$       & Total transfer rounds & \REMOP \\
    $L$       & Total latency cost (seconds) & \REMOP \\
    $\tau$    & Network parameter ($\propto BW\cdot RTT$) & \REMOP \\
    $\alpha$    & Memory-scaled network parameter ($M/\tau$) & \REMOP \\
    $R_{in}$, $R_{out}$  & Input/output buffer size & \REMOP \\
    $r_{in}$, $r_{out}$  & Input/output buffer ratio ($R_{in}/M$, $R_{out}/M$) & \REMOP \\
    $|R|$, $|S|$ & Size of NL join relations $R$, $S$ & NL Join \\
    $P_R$, $P_S$ & Budget in input buffer for $R$ and $S$  & NL Join \\
    $p_R$, $p_S$ & Fraction of budget for $R$ and $S$ & NL Join \\
    $O$       & Join output size & NL Join \\
    $f$       & Join output selectivity ($O/(|R||S|)$) ($\text{page}^{-1}$) & NL Join \\
    $\beta$       & Selectivity--memory parameter ($fM$) & NL Join \\
    $\ell$    & Number of sorted runs & Sort \\
    $k$       & Merge fan-in & Sort \\
    $|B|$, $|Q|$  & Size of build/probe relations & Hash Join \\
    $P$       & Number of hash partitions & Hash Join \\
    $\sigma$   & Fraction of spilled partitions & Hash Join \\
    $M_{HT}$, $M_b$ & Hash-table and I/O buffer-pool size & Hash Join \\
    $R_{r\text{/}w\text{/}s\text{/}o}$ & Per-phase read/write/stage/output buffers & Hash Join \\
    \bottomrule
  \end{tabular}
\end{table}

\subsection{Case Studies: BNLJ and EMS}
\label{subsec:example}

\begin{figure}[t]
  \centering
  \begin{minipage}{0.4\linewidth}
    \centering
    \includegraphics[width=\linewidth]{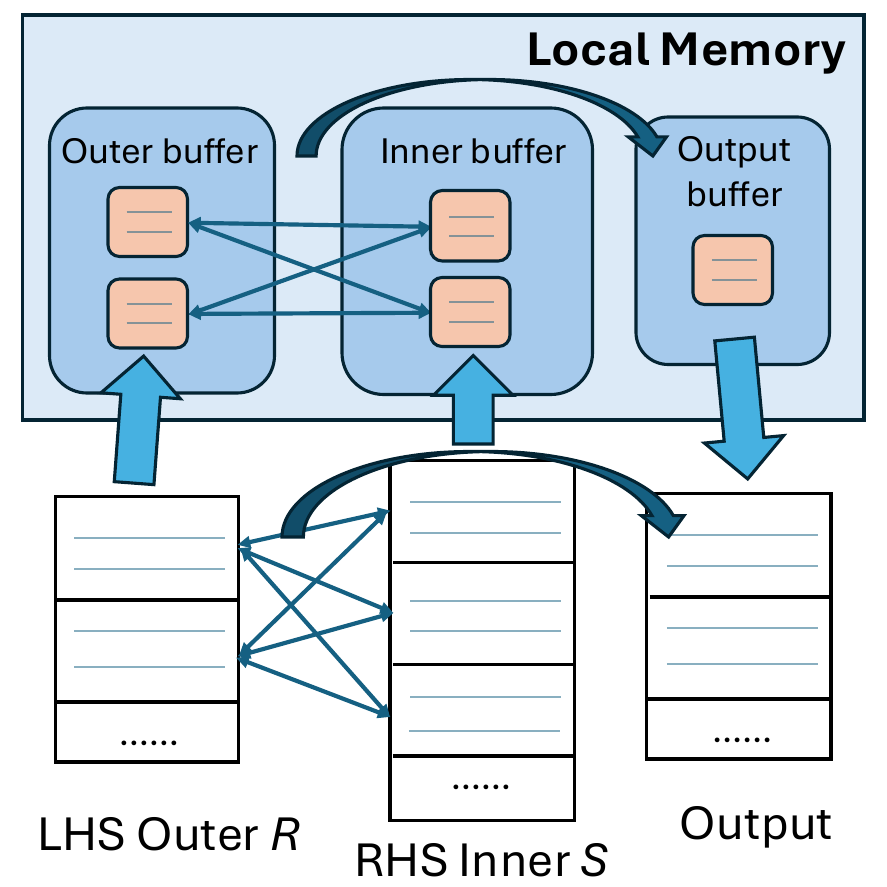}
  \end{minipage}\hfill
  \begin{minipage}{0.55\linewidth}
    \centering
    \includegraphics[width=\linewidth]{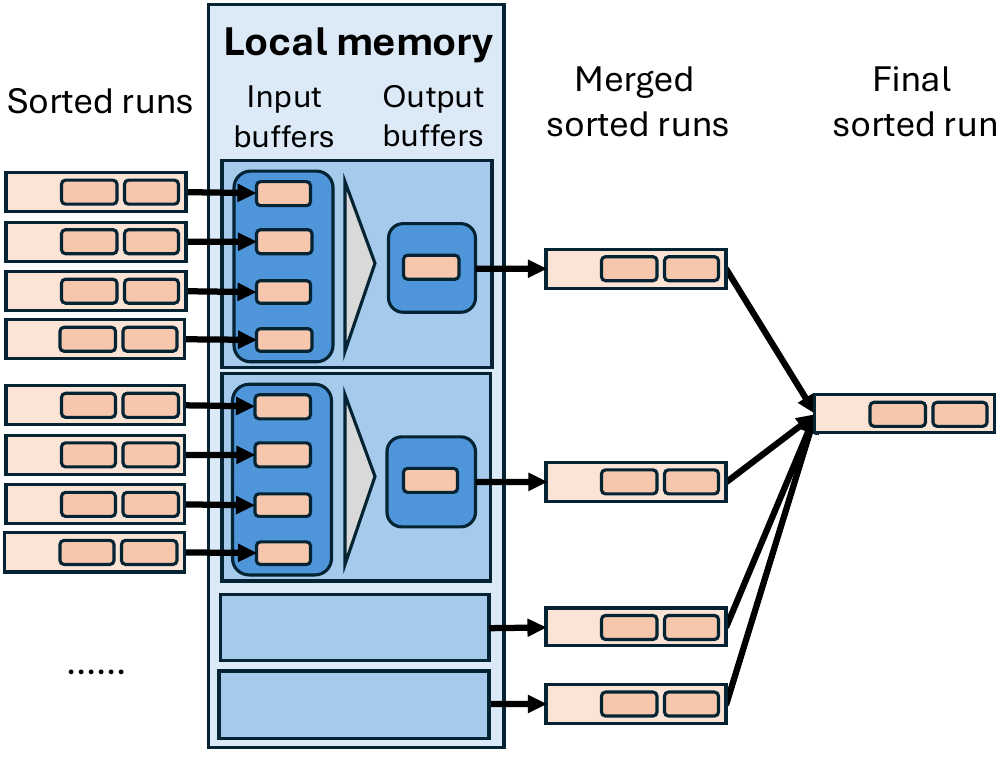}
  \end{minipage}
  \caption{Out-of-memory operator execution. Left: Blocked nested-loop join. Right: $k$-way merge sort.}
  \Description{Two-panel illustration of out-of-memory operator execution: (left) a blocked nested-loop join where the input region is split between the outer and inner relations and the operator scans the inner relation in blocks for each outer block; (right) a $k$-way merge sort merge phase where the input region is partitioned across $k$ run buffers and the output is periodically flushed.}
  \label{fig:nlj_ems}
  \label{fig:merge_example}
\end{figure}

We illustrate \REMOP's optimization opportunities through concrete examples of two common analytical operators: \emph{blocked nested-loop join} (BNLJ) and \emph{external merge sort} (EMS).
They represent repeated input scans and multi-pass streaming, respectively, and show how minimizing $D$ can lead to a substantially different allocation from minimizing $C$.
Figure~\ref{fig:nlj_ems} shows their execution patterns, and the following examples quantify this trade-off.

\paragraph{Blocked Nested-Loop Join}
In BNLJ, when both relations exceed $M$, each memory-resident block of the outer relation $R$ triggers a full scan of the inner relation $S$, effectively performing an all-to-all comparison (Figure~\ref{fig:nlj_ems}, left).

Suppose that the two relations have $|R|{=}500$, $|S|{=}1{,}000$, and $M{=}101$ pages.
We reserve one page for output and compare the read-side costs.
The conventional allocation~\cite{Shapiro1986Joins} places nearly all memory on the outer $R$: $P_R{=}99$, $P_S{=}1$, and $R_{out}{=}1$.
It reads
$D_{read}{=}\lceil |R|/P_R\rceil|S|{+}|R|{=}\lceil 500/99 \rceil {\times} 1{,}000{+}500{=}6{,}500$ pages in
$C_{read}{=}\lceil |R|/P_R\rceil\lceil |S|/P_S\rceil{+}\lceil |R|/P_R\rceil{=}6{,}006$ rounds.

In contrast, with an equal split ($P_R{=}P_S{=}50$, $R_{out}{=}1$),
$D_{read}{=}\lceil500/50\rceil {\times} 1{,}000{+}500{=}10{,}500$ pages, while
$C_{read}{=}\lceil 500/50 \rceil {\times} \lceil 1{,}000/50 \rceil{+}\lceil500/50\rceil{=}210$ rounds.
Therefore, the equal split transfers $61.5\%$ more data but uses $96.5\%$ fewer rounds, which can significantly reduce latency cost in remote memory.

\paragraph{External Merge Sort}
EMS partitions the input into multiple runs, sorts each in memory, and merges them over one or more passes.
We focus on the merge phase, which merges $\ell$ sorted runs with fan-in $k$ by repeatedly refilling per-run input buffers and flushing an output buffer (Figure~\ref{fig:nlj_ems}, right).

Given $N{=}13{,}000$ pages of data in $\ell{=}128$ sorted runs, and $M{=}101$ pages, the conventional choice maximizes fan-in with $k{=}M{-}1$, one page per run, and one output page~\cite{Graefe2006Sorting}.
This runs $\lceil \log_{k} \ell \rceil {=} 2$ merge passes, yielding $D{=}D_{read}{+}D_{write}{=}2{\times} 2N{=}52{,}000$ pages and $C{=}C_{read}{+}C_{write}{=}2(N{+}N){=}52{,}000$ rounds.

However, if we choose $k{=}4$ with an input-to-output ratio of $2{:}1$ (i.e., $R_{in}{=}67$, $R_{out}{=}34$), it runs $\lceil \log_{4} 128 \rceil{=}4$ merge passes, yielding $D{=}D_{read}{+}D_{write}{=}2 {\times} 4 N {=} 104{,}000$ pages and $C{=}C_{read}{+}C_{write}{=}4 (\lceil N / \lfloor R_{in} / k \rfloor \rceil {+} \lceil N / R_{out} \rceil) {=} 4{,}784$.
This doubles $D$ but reduces $C$ by roughly $10\times$.

Overall, these examples show that tuning internal buffer allocations can substantially reduce transfer rounds while accepting additional data transfer, a favorable trade-off when remote-memory round-trip latency dominates over bandwidth.


\section{Operator Algorithms and Analysis}
\label{sec:algo}

In this section, we instantiate \REMOP's optimization framework (\S\ref{sec:fm}) on three representative operators: \emph{blocked nested-loop join} (BNLJ), \emph{external merge sort} (EMS), and \emph{external hash join} (EHJ).
For each operator, we parameterize its buffer allocation, derive its $C$, $D$, and $L$-optimal policy, and compare it with conventional and DuckDB strategies.

\subsection{\texorpdfstring{Blocked Nested-Loop Join}{Blocked Nested-Loop Join}}
\label{subsec:nlj}

\paragraph{Problem Setup} 
Consider joining outer relation $R$ and inner relation $S$ under predicate $\Phi$ and budget $M$. The input region $R_{in}$ is split into $P_R=p_RR_{in}$ and $P_S=p_SR_{in}$ pages, where $p_R+p_S=1$, while the remaining $R_{out}=M-R_{in}$ pages buffer output.

BNLJ (Algorithm~\ref{alg:bnlj}) loads each $P_R$-page outer block and scans $S$ in $P_S$-page blocks, buffering matches in the $R_{out}$-page output region. Each block read or output flush is one transfer round, so enlarging one region reduces its rounds but leaves less memory for the others.

\begin{algorithm}[t]
  \caption{Blocked Nested-Loop Join}
  \label{alg:bnlj}
  \begin{algorithmic}[1]
    \Require{$R$, $S$, predicate $\Phi$; budget $M$, ratios $r_{in}$ and $p_R$}
    \State $R_{in} \gets r_{in} \cdot M$; \quad $R_{out} \gets M - R_{in}$
    \State $P_R \gets p_R \cdot R_{in}$; \quad $P_S \gets (1 - p_R) \cdot R_{in}$
    \ForAll{blocks $R_i$ of $R$ with size $P_R$} \\\Comment{$\lceil|R|/P_R\rceil$ blocks}
      \State \textbf{Read} $R_i$ into buffer \Comment{1 read round}
      \ForAll{blocks $S_j$ of $S$ with size $P_S$}\\\Comment{$\lceil|S|/P_S\rceil$ blocks}
        \State \textbf{Read} $S_j$ into buffer \Comment{1 read round}
        \State Join $R_i\times S_j$; buffer matches
        \State \textbf{Flush} output buffer when full
        \Comment{1 write round}
      \EndFor
    \EndFor
  \end{algorithmic}
\end{algorithm}

\paragraph{Cost Analysis}
Each outer block triggers one read and a blocked scan of $S$, giving approximately $(|R||S|)/(P_RP_S)$ read rounds, while flushing $O$ output pages adds $O/R_{out}$ rounds. Moreover, each outer block is read once, but $S$ is rescanned once per outer block. Thus,
\[
\begin{aligned}
C_{\text{BNLJ}}&\approx\frac{|R||S|}{p_Rp_SR_{in}^{2}}+\frac{O}{R_{out}},\\
D_{\text{BNLJ}}&=|R|+\left\lceil\frac{|R|}{P_R}\right\rceil|S|+O
\approx |R|+\frac{|R||S|}{p_RR_{in}}+O .
\end{aligned}
\]
We optimize the two allocation knobs, $p_R{:}p_S$ and $r_{in}=R_{in}/M$, under $L_{\text{BNLJ}}=D_{\text{BNLJ}}+\tau C_{\text{BNLJ}}$.

\paragraph{Optimal Inner/Outer Split ($p_R : p_S$)}
Intuitively, the split trades volume against rounds: a larger $p_R$ shrinks $D_{\text{BNLJ}}$ (fewer inner rescans) but inflates $C_{\text{BNLJ}}$. Minimizing $L_{\text{BNLJ}}$ strikes a balance in this trade-off.

\begin{property}
\label{lem:nlj_split}
For a fixed input region $R_{in}$, the latency cost $L_{\text{BNLJ}}$ is minimized at the split
\[
p_R^\star : p_S^\star = \sqrt{1 + R_{in}/\tau}:1, \qquad p_R^\star + p_S^\star = 1.
\]
\end{property}

\begin{proof}[Proof]
After removing constants, the split-dependent objective is $1/p_R+\tau/[R_{in}p_R(1-p_R)]$. It is strictly convex on $(0,1)$, and setting its derivative to zero gives $p_R^\star/p_S^\star=\sqrt{1+R_{in}/\tau}$.
\end{proof}

The solution connects two regimes. As $\tau\to\infty$, it approaches a size-independent $1{:}1$ equal split that minimizes the transfer round cost. As $\tau\to0$, $p_S^\star\to0$, recovering the outer-heavy allocation that minimizes the data transfer cost.

\paragraph{Optimal Input/Output Ratio ($r_{in} : r_{out}$)}
Let the \emph{output selectivity} $f=O/(|R||S|)$ be the fraction of the Cartesian product emitted~\cite{Selinger1979}. Given $p_R^\star$ derived in Property~\ref{lem:nlj_split}, the terms in $L_{\text{BNLJ}}$ affected by $r_{in}$ are
\[
  \min_{r_{in}\in(0,1)}\; g(r_{in}) = \frac{1}{p_R^\star \cdot r_{in}} + \frac{1}{p_R^\star (1-p_R^\star)} \cdot \frac{\tau}{M r_{in}^2} + \frac{f \cdot \tau}{1 - r_{in}}.
\]
The optimum depends on $\alpha=M/\tau$ and $\beta=fM$, as Table~\ref{tab:rin_vs_f} shows. In practice, PK--FK equijoins\footnote{A PK--FK equijoin joins a primary-key column to a foreign-key column; typically each FK tuple matches at most one PK tuple.} typically have $f\sim1/n_P\in[10^{-7},10^{-2}]$~\cite{Selinger1979, Leis2015HowGood}, hence $\beta\ll1$ and $r_{in}^\star\approx1$. 
Low-cardinality many-to-many joins can reach $f\sim10^{-2}$--$10^{-1}$~\cite{Selinger1979}, requiring more output buffering.
\begin{table}[t]
  \centering
  \caption{Optimal input ratio $r_{in}^{\star}(\alpha,\beta)$ for BNLJ.}
  \label{tab:rin_vs_f}
  \begin{tabular}{c|ccccccc}
    \toprule
    $\beta \backslash \alpha$ & $10^{-2}$ & $10^{-1}$ & $1$ & $10$ & $10^{2}$ & $10^{3}$ & $10^{4}$ \\
    \midrule
    $10^{-2}$ & 0.966 & 0.967 & 0.970 & 0.980 & 0.991 & 0.997 & 0.999 \\
    $10^{-1}$ & 0.904 & 0.905 & 0.912 & 0.940 & 0.973 & 0.991 & 0.997 \\
    $1$       & 0.764 & 0.765 & 0.778 & 0.836 & 0.921 & 0.971 & 0.990 \\
    $10$      & 0.547 & 0.549 & 0.560 & 0.633 & 0.789 & 0.913 & 0.970 \\
    $10^{2}$  & 0.330 & 0.331 & 0.337 & 0.384 & 0.549 & 0.769 & 0.910 \\
    \bottomrule
  \end{tabular}
\end{table}

\paragraph{Comparison with Default Strategies}
Disk-based BNLJ assigns $P_R=M-2$, $P_S=1$, and $R_{out}=1$, yielding
$C_{\text{BNLJ}}^{\text{conv}}\approx|R||S|/(M-2)+O$ and
$D_{\text{BNLJ}}^{\text{conv}}=|R|+\lceil|R|/(M-2)\rceil|S|+O$.
Against an equal input split, this incurs up to $R_{in}^{2}/(4M)$ times as many read rounds and $R_{out}$ times as many write rounds. For the dominant inner scans, the equal split sets $P_R=R_{in}/2$ instead of $P_R\approx M$, increasing the data transfer cost by $M/(R_{in}/2)=2/r_{in}$. 
Thus, the conventional allocation minimizes transferred data, whereas \REMOP accepts bounded additional transfer to reduce RTT-dominated rounds.

\subsection{\texorpdfstring{$k$}{k}-Way Merge Sort}
\label{subsec:kway}

\paragraph{Problem Setup}
External merge sort is a common algorithm for sorting data that exceeds memory capacity~\cite{Graefe2006Sorting, AggarwalVitter1988IOComplexity, PangCareyLivny1993MemoryAdaptiveSorting}.
It consists of two phases: an in-memory sorting phase that produces $\ell = \lceil N / M \rceil$ sorted runs, and a merge phase that merges these runs into a single sorted output using a $k$-way merge algorithm.

During merging, $R_{in}$ pages are shared by the $k$ input runs and $R_{out}=M-R_{in}$ pages buffer output. Both $k$ and $r_{in}=R_{in}/M$ are tunable parameters.

The EMS merge phase (Algorithm~\ref{alg:kway}) merges groups of $k$ remote runs, assigning about $R_{in}/k$ pages to each input and $R_{out}$ pages to output.
In each group of $k$ runs, a selection structure (e.g., a min-heap or tournament tree) chooses the smallest current run head and appends it to the output. 
The operator refills an input buffer when it empties and flushes the output buffer when full, with each refill or flush forming a batched transfer round. 
Each pass reduces the run count by a factor of $k$, requiring $\lceil\log_k\lceil N/M\rceil\rceil$ passes. 
Thus, $k$ and the buffer split jointly determine rounds per pass and pass count.

\begin{algorithm}[t]
  \caption{$k$-Way Merge Sort}
  \label{alg:kway}
  \begin{algorithmic}[1]
    \Require{$\ell$ sorted runs of $N$ pages; $M$, fan-in $k$, ratio $r_{in}$}
    \State $R_{in}\gets r_{in}M$; \quad $R_{out}\gets M-R_{in}$
    \While{$\ell>1$}
      \ForAll{groups of $k$ runs}
        \State Allocate $R_{in}/k$ per input and $R_{out}$ for output
        \While{not all $k$ runs exhausted}
          \State Merge current heads into output
          \State \textbf{Refill} empty inputs
          \Comment{1 read round} 
          \State \textbf{Flush} full output blocks
          \Comment{1 write round}
        \EndWhile
      \EndFor
      \State $\ell\gets\lceil\ell/k\rceil$ \Comment{1 merge pass}
    \EndWhile
  \end{algorithmic}
\end{algorithm}

\paragraph{Cost Analysis}
In one pass, each of the $\ell$ runs is read through an $R_{in}/k$-page buffer, giving approximately $kN/R_{in}$ read rounds; output adds $N/R_{out}$ write rounds. Every page is read and written once, so $D_{\text{pass}}=2N$. Multiplying by $\log_k(N/M)$ passes gives
\[
L_{\text{EMS}} \approx \left(2 + \tau \cdot \left(\frac{k}{R_{in}} + \frac{1}{R_{out}}\right) \right) \cdot N \log_k \left(\frac{N}{M}\right)
\]

\paragraph{Optimal Input/Output Ratio ($r_{in} : r_{out}$)}
For fixed $k$, the split affects only $C$ because $D_{\text{pass}}=2N$. The optimum of $L_{\text{EMS}}$ is therefore independent of $\tau$.

\begin{property}
\label{lem:kway_ratio}
For $k$-way merge sort, the transfer-round cost is minimized when $R_{in}:R_{out}=\sqrt{k}:1$.
\end{property}

\begin{proof}[Proof]
Minimizing $k/R_{in}+1/(M-R_{in})$ gives $R_{in}/R_{out}=\sqrt{k}$.
\end{proof}

\paragraph{Optimal Fan-In ($k$)}
Substituting $R_{in}=\sqrt{k}M/(\sqrt{k}+1)$ and $R_{out}=M/(\sqrt{k}+1)$ yields
\[
L_{\text{EMS}}(k) = N\log_2\!\frac{N}{M}\cdot\frac{\,2 + \tau(\sqrt{k}+1)^2/M\,}{\log_2 k}.
\]
With $\alpha=M/\tau$, minimizing the objective $h(k)=[2+(\sqrt{k}+1)^2/\alpha]/\log_2k$ gives Table~\ref{tab:kstar}.
The RTT-dominated limit is $k^\star=4$; as $\tau$ falls, $k^\star$ grows toward the largest feasible fan-in.

\begin{table}[t]
  \centering
  \caption{Optimal fan-in $k^{\star}(\alpha)$ for EMS ($\alpha=M/\tau$).}
  \label{tab:kstar}
  \begin{tabular}{c|ccccccc}
    \toprule
    $\alpha$ & $(\to 0)$ & $1$ & $4$ & $16$ & $64$ & $256$ & $1024$ \\
    \midrule
    $k^{\star}$ & 4 & 5 & 8 & 17 & 43 & 126 & 396 \\
    \bottomrule
  \end{tabular}
\end{table}

\paragraph{Comparison with Default Strategies}
Conventional merging minimizes passes by setting $k=M-1$ with one page per input and output. This minimizes $D$ but incurs about $2N$ transfer rounds per pass, giving
$L^{\text{conv}}_{\text{EMS}}=2N(1+\tau)\log_{M-1}(N/M)$.
Its extreme fan-in is also complex to implement in practice and costly at large $\tau$.

DuckDB (v1.0.0)~\cite{DuckDB2019} uses 2-way merge with one block per input and output, equivalent to $k=2$, $R_{in}=2M/3$, and $R_{out}=M/3$. Thus,
$L^{\text{Duck}}_{\text{EMS}}=2N(1+3\tau/M)\log_2(N/M)$.
As $\tau\to\infty$, conventional merging and DuckDB are respectively about $4M/(9\log_2M)$ and $4/3$ times higher than the optimal 4-way merge in terms of transfer rounds.

\subsection{External Hash Join}
\label{subsec:hj}

\paragraph{Problem Setup}
Consider hash-joining build relation $B$ and probe relation $Q$ under budget $M$. Both are hash-partitioned into $P=2^b$ partitions. Resident hash tables occupy $M_{HT}$ pages, while the phases reuse an I/O pool $M_b=M-M_{HT}$. A fraction $\sigma$ of the partitions spill and the rest remain resident.

EHJ (Algorithm~\ref{alg:hj}) consists of three phases: \emph{Build phase}: partition $B$ while building resident hash tables and spilling the rest; \emph{Probe phase}: partition $Q$ while probing resident tables and staging spilled tuples; \emph{External phase}: join the spilled partition pairs in rounds.

\begin{algorithm}[t]
  \caption{External Hash Join}
  \label{alg:hj}
  \begin{algorithmic}[1]
    \Require{Build $B$, probe $Q$; budget $M$, radix partition $P$}
    \State \textit{// P1: partition build} \Comment{read $R_r$, write $R_w$}
    \ForAll{blocks $B_i$ of $B$}
      \State \textbf{Read} block $B_i$ into buffer \Comment{1 read round}
      \State Build resident tuples and buffer spilled tuples
      \State \textbf{Flush} write buffers when full \Comment{1 write round}
    \EndFor
    \State \textit{// P2: partition probe} \Comment{read $R_r$, stage $R_s$, output $R_o$}
    \ForAll{blocks $Q_i$ of $Q$}
      \State \textbf{Read} block $Q_i$ into buffer \Comment{1 read round}
      \State Probe resident and stage spilled tuples
      \State \textbf{Flush} stage/output buffers when full \Comment{1 write round}
    \EndFor
    \State \textit{// P3: external rounds} \Comment{read $R_r$, output $R_o$}
    \ForAll{spilled partition pairs}
      \State \textbf{Read} and join the pair
      \Comment{1 read round}
      \State \textbf{Flush} output buffers when full
      \Comment{1 write round}
    \EndFor
  \end{algorithmic}
\end{algorithm}

\paragraph{Per-Phase Cost Analysis}
Table~\ref{tab:hj_cost} collects each phase's costs. In P1, reading the build side $B$ contributes $|B|$ pages to $D_1$ and $|B|/R_r$ rounds to $C_1$. For spilled partitions, it adds $\sigma|B|$ pages to $D_1$, and since $R_w$ is divided into $\sigma P$ partition buffers of size $R_w/(\sigma P)$, it requires $\sigma|B|/[R_w/(\sigma P)]=\sigma^2P|B|/R_w$ spilling rounds. P2 analogously reads $Q$, stages $\sigma|Q|$ pages through $R_s$, and writes $(1-\sigma)O$ resident output through $R_o$, yielding the three terms in $C_2$. Finally, P3 rereads $\sigma(|B|+|Q|)$ pages through $R_r$ and writes the remaining $\sigma O$ output through $R_o$.

\begin{table}[t]
  \centering
  \caption{Per-phase costs for EHJ.}
  \label{tab:hj_cost}
  \small
  \resizebox{\ifdim\width>\columnwidth\columnwidth\else\width\fi}{!}{%
  \begin{tabular}{c|c|c}
    \toprule
    Phase & data transfer cost $D_i$ & transfer round cost $C_i$ \\
    \midrule
    P1 (build) & $(1+\sigma)|B|$ & $\frac{|B|}{R_r}+\frac{\sigma^{2}P|B|}{R_w}$ \\[4pt]
    P2 (probe) & $(1+\sigma)|Q|+(1-\sigma)O$ & $\frac{|Q|}{R_r}+\frac{\sigma^{2}P|Q|}{R_s}+\frac{(1-\sigma)O}{R_o}$ \\[4pt]
    P3 (ext.)  & $\sigma(|B|+|Q|)+\sigma O$ & $\frac{\sigma(|B|+|Q|)}{R_r}+\frac{\sigma O}{R_o}$ \\
    \bottomrule
  \end{tabular}}
\end{table}

\paragraph{Optimal Allocation}
For fixed $|B|$, $|Q|$, and $\sigma$, buffer allocation does not affect $D_i$. Thus, minimizing $L_i=D_i+\tau C_i$ reduces to minimizing $C_i$.

\begin{property}
\label{lem:hj_split}
For each phase, the buffer split that minimizes $C_i$ and the resulting minimum $C_i^{\star}$ are listed in Table~\ref{tab:hj_opt}.
\end{property}

\begin{proof}[Proof]
Each $C_i$ has the form $\sum_j a_j/R_j$ with coefficients $a_j$ and $\sum_jR_j=M_b$. The Cauchy--Schwarz inequality gives the minimum $C_i^{\star}=\big(\sum_j\sqrt{a_j}\big)^{2}/M_b$ at $R_j\propto\sqrt{a_j}$; substituting the phase coefficients yields Table~\ref{tab:hj_opt}.
\end{proof}

\begin{table}[t]
  \centering
  \caption{Per-phase optimal buffer split for EHJ.}
  \label{tab:hj_opt}
  \small
  \resizebox{\ifdim\width>\columnwidth\columnwidth\else\width\fi}{!}{%
  \begin{tabular}{c|c|c}
    \toprule
    Phase & optimal split & transfer-round cost $C_i^{\star}$ \\
    \midrule
    P1 & $R_r{:}R_w=1{:}\sigma\sqrt{P}$ & $\frac{|B|(1+\sigma\sqrt{P})^{2}}{M_b}$ \\[4pt]
    P2 & $R_r{:}R_s{:}R_o=1{:}\sigma\sqrt{P}{:}\sqrt{\frac{(1-\sigma)O}{|Q|}}$ & $\frac{\big(\sqrt{|Q|}+\sigma\sqrt{P|Q|}+\sqrt{(1-\sigma)O}\big)^{2}}{M_b}$ \\[4pt]
    P3 & $R_r{:}R_o=\sqrt{|B|+|Q|}{:}\sqrt{O}$ & $\frac{\sigma\big(\sqrt{|B|+|Q|}+\sqrt{O}\big)^{2}}{M_b}$ \\
    \bottomrule
  \end{tabular}}
\end{table}

\paragraph{Comparison with the Default Allocation}
DuckDB fixes $P=16$ and repartitions oversized partitions. It provides a default-sized probe-spill buffer and no dedicated write pool. Under memory pressure, per-block evictions make the $\sigma^2P/R$ write-round terms large. \REMOP keeps DuckDB's partitioning but enlarges $R_w$ and $R_s$ toward Property~\ref{lem:hj_split}'s splits, using buffer memory to reduce write rounds.

\subsection{Summary}
\label{subsec:summary}

Table~\ref{tab:comparison} summarizes the derived policies. 
For BNLJ, $p_R^\star{:}p_S^\star=\sqrt{1+R_{in}/\tau}{:}1$ approaches an equal split as $\tau\to\infty$; compared with the outer-heavy policy, it reads up to $2/r_{in}$ times more data but cuts rounds by $\Theta(M)$. 
For EMS, $R_{in}{:}R_{out}=\sqrt{k}{:}1$ and RTT-dominated execution selects $k^\star=4$, which uses $25\%$ fewer rounds than DuckDB's 2-way merge. 
For EHJ, each phase allocates $R_j\propto\sqrt{a_j}$ for its load coefficient $a_j$. Since this changes $C_i$ but not $D_i$, its structural gain is smaller than for BNLJ and EMS.

Across the operators, \REMOP minimizes $L=D+\tau C$ rather than $D$ alone. The $\tau\to0$ limit recovers conventional data-minimizing policies, while larger $\tau$ favors fewer, larger transfers.
More broadly, \REMOP turns remote-memory-aware operator design into a principled buffer-allocation problem. It jointly accounts for transferred data and round trips, derives deployment-aware policies, and applies to out-of-memory operators built from repeated batched reads and flushes. Section~\ref{sec:impl} implements these policies in DuckDB.

\begin{table}[t]
  \centering
  \caption{Buffer-allocation ratios under each strategy.}
  \label{tab:comparison}
  \begin{tabular}{llcc}
    \toprule
    Op. & Strategy & $R_{in}$:$R_{out}$ & $P_R$:$P_S$ / $k$ / $P$ \\
    \midrule
    \multirow{2}{*}{BNLJ} & Conv. & $(M{-}1)$:$1$ & $(M{-}2)$:$1$ \\[3pt]
    & \REMOP & $r_{in}^{\star}$ (Tab.~\ref{tab:rin_vs_f}) & $\sqrt{1{+}R_{in}/\tau}{:}1$ \\[3pt]
    \midrule
    \multirow{3}{*}{EMS} & Conv. & $(M{-}1)$:$1$ & $M{-}1$ \\[3pt]
    & DuckDB & $2:1$ & $2$ \\[3pt]
    & \REMOP & $\sqrt{k}{:}1$ & $k^{\star}$ (Tab.~\ref{tab:kstar}) \\[3pt]
    \midrule
    \multirow{2}{*}{EHJ} & DuckDB & default & $16$ \\[3pt]
    & \REMOP & $R_j^{\star}$ (Tab.~\ref{tab:hj_opt}) & $16$ \\[3pt]
    \bottomrule
  \end{tabular}
\end{table}


\section{Implementation}
\label{sec:impl}

We implement \REMOP in DuckDB~\cite{DuckDB2019} (v1.0.0). This section presents its architecture (\S\ref{subsec:sys_overview}), three optimized operators (\S\ref{subsec:nlj_impl}, \S\ref{subsec:kway_impl}, \S\ref{subsec:hj_impl}), prefetch buffer (\S\ref{subsec:prefetch_impl}), and remote-memory integration (\S\ref{subsec:itgr}).

\subsection{System Overview}
\label{subsec:sys_overview}

\begin{figure}[t]
  \centering
  \includegraphics[width=0.8\linewidth]{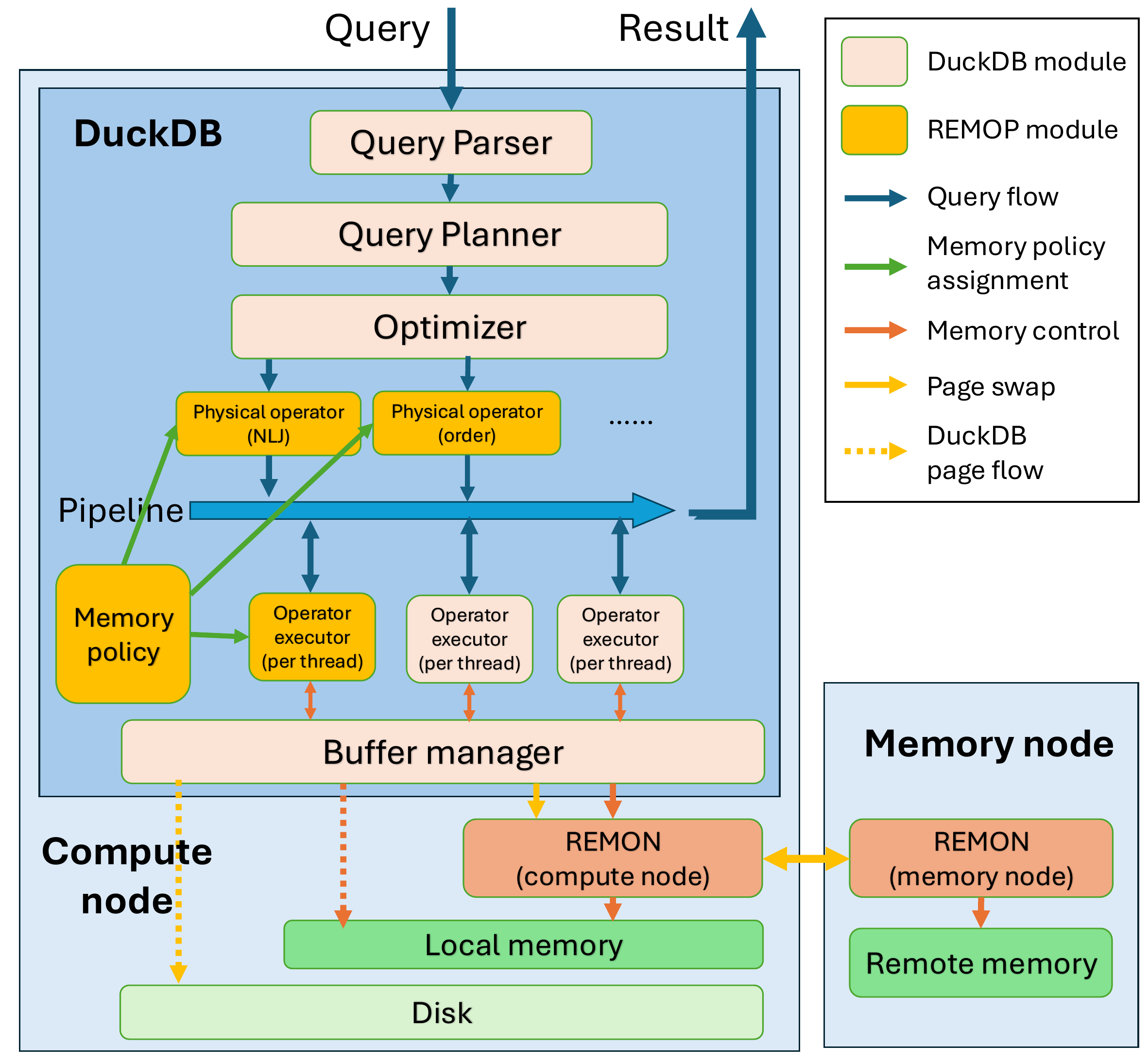}
  \caption{\REMOP system architecture: modified DuckDB with remote-memory integration.}
  \Description{System architecture of \REMOP in DuckDB, including planner, memory policy module, operators, buffer manager, and remote-memory backend.}
  \label{fig:architecture}
\end{figure}

Figure~\ref{fig:architecture} shows the DuckDB architecture with \REMOP integrated. We implement our tunable memory-allocation strategies as new operator variants in DuckDB's pipeline engine. A \emph{memory policy module} between the query planner and these physical operators assigns per-operator budgets and the allocation parameters derived in \S\ref{sec:algo}:
\begin{itemize}[leftmargin=*]
  \item \textbf{BNLJ:} memory budget~$M$, input ratio~$r_{in}$, and outer fraction~$p_R$.
  \item \textbf{EMS:} memory budget~$M$, input ratio~$r_{in}$, and fan-in~$k$.
  \item \textbf{EHJ:} 
  memory budget~$M$, radix~$P$, build write pool~$R_w$, and probe stage pool~$R_s$.
\end{itemize}
These parameters are exposed as DuckDB runtime settings,
allowing the optimizer or the user to tune them at query time.
When the master switch \texttt{operator\_memory\_optimizations} is enabled, the parameters are locked to the configured values.
At execution time, each operator executor reads its policy and configures its internal buffers accordingly.

Under the operator executors, DuckDB's buffer manager mediates page access and evicts pages to remote memory when local DRAM is exhausted. Operators realize their policies through execution logic and \emph{pin/unpin} calls, while the buffer manager invokes the remote-memory system's evict/fetch interfaces for page movement.

\noindent\textbf{Multi-threading:} To support parallel execution, we divide the memory budget equally across threads.
Each executor configures its buffers for the per-thread budget and executes tasks scheduled by the query planner and physical operators.

\subsection{BNLJ Implementation}
\label{subsec:nlj_impl}

DuckDB's standard NLJ materializes the shorter RHS (inner) relation $S$, then streams LHS (outer) relation $R$ chunks and compares each against all RHS chunks, which is costly under remote memory.

Our BNLJ extends DuckDB's NLJ with configurable buffer allocation. 
It falls back to the standard in-memory path when the inner relation fits its budget.
Otherwise, it performs the buffer-oriented execution with a four-phase state machine:
\textbf{(1) Accumulate:} incoming LHS outer-relation ($R$) chunks are accumulated in a buffer-managed collection of up to $P_R$ pages.
\textbf{(2) Process blocks:} for each buffered outer block, the operator scans $S$ in $P_S$-page blocks, joins each block pair in memory, and appends matches to the $R_{out}$-page output buffer.
\textbf{(3) Drain results:} when the output buffer is full, it is flushed downstream before execution resumes from the paused position.
\textbf{(4) Finished:} all outer blocks have been processed and all results drained.

\noindent\textbf{Block pinning:}
As in Algorithm~\ref{alg:bnlj}, each outer block remains pinned while the operator scans inner blocks, which are pinned and unpinned one at a time. Results are accumulated until the output buffer fills, reducing pipeline interruptions and amortizing remote-write round overhead.

\noindent\textbf{Blockwise variant:}
For non-decomposable predicates such as theta joins, we provide a blockwise variant that evaluates the predicate on the cross-product of each block pair while sharing BNLJ's state machine, allocation, and result buffering.

\subsection{EMS Implementation}
\label{subsec:kway_impl}

Our $k$-way EMS replaces DuckDB's default two-way merge and is structured as a multi-pass merge pipeline.

\noindent\textbf{Merge passes:}
Each pass groups up to $k$ runs produced by the initial sort or previous pass. For each group, the sorter distributes $R_{in}$ pages across the runs (about $R_{in}/k$ each), reserves $R_{out}$ pages for output, and uses a tournament tree to select and advance the run with the smallest key. Exhausted runs are marked inactive, and the resulting runs enter the next pass until one final run remains.

\noindent\textbf{Batched merge pipeline:}
Instead of copying one tuple at a time, the merge pipeline operates in four batched phases: 
(1)~pop a batch of tournament-tree winners, 
(2)~copy radix-sort keys from their source blocks, 
(3)~copy variable-length blob keys, and 
(4)~copy payloads. 
Grouping copies from the same source block improves cache locality and reduces pin/unpin operations. 
When only one run remains active, the sorter bypasses the tournament tree and bulk-copies its blocks.

\noindent\textbf{Parallel merge:}
To parallelize a merge group, a partitioner divides the merged key range into non-overlapping partitions. Each partition contains one contiguous section from every run; the partitioner finds their boundaries by binary search and targets a total size of about $R_{in}$ pages. Threads then merge assigned partitions concurrently through private pipeline instances and output buffers, after which the results are concatenated in order.

\subsection{EHJ Implementation}
\label{subsec:hj_impl}

DuckDB executes large equi-joins with an external radix-partitioned hash join that splits both sides into $P$ partitions. It spills a fraction of the partitions and joins each spilled pair in later rounds. This realizes the build, probe, and external phases analyzed in \S\ref{subsec:hj}.

\REMOP preserves this execution and configures radix~$P$ and the two write-side pools optimized by our implementation. It realizes $R_w$ through a minimum number of pinned page handles per build partition and $R_s$ through a staging capacity in rows per spilled probe partition.

\noindent\textbf{Build:} The operator reads the build side through the $R_r$-page input buffer and partitions it at radix~$P$. Resident partitions are built into hash tables in place, while spilled partitions are flushed through the $R_w$ write pool. The pinned-handle knob retains a minimum number of pages for each build partition before releasing them, batching partition writes and realizing the $R_w$ allocation.

\noindent\textbf{Probe:} The probe side is partitioned to match the build side. Tuples in resident partitions probe their hash tables on the fly, while spilled tuples accumulate before being flushed. The staging-row knob sets each spilled probe partition's row capacity, where a larger value batches more tuples per flush.

\noindent\textbf{External rounds:} Each spilled build partition is reloaded into a hash table and joined with its matching probe partition.

The resident hash-table reservation and the spill fraction~$\sigma$ are governed by DuckDB's temporary-memory manager rather than by the memory policy module. Therefore, \REMOP tunes the per-phase buffer pools around a system-determined~$\sigma$.

\subsection{Prefetch Buffer}
\label{subsec:prefetch_impl}

The allocation strategies above reduce transfer rounds, yet each remaining round still stalls the operator thread for about one RTT when fetching a cold page.
As a complementary, operator-agnostic optimization, we add a prefetch double buffer that overlaps the next input fetch with the processing of the current batch. While the operator consumes one slot, a background worker fills the other through the buffer manager, largely hiding the fetch latency when batch processing lasts longer than one RTT.

We place the buffer at one input-consuming site per operator: BNLJ prefetches inner chunks scanned against each resident outer block, EMS prefetches a run's next input block as its current block drains, and EHJ prefetches spilled probe partitions in external rounds. Each operator independently issues prefetching when the current batch is half consumed.
The extra slot comes from its input region, preserving budget~$M$.

\subsection{Integration with Remote Memory}
\label{subsec:itgr}

\REMOP's DuckDB implementation is backend-agnostic, as long as the underlying memory system exposes page-level memory controls.
To demonstrate this generality, we integrate DuckDB with two remote-memory backends that use different network stacks: \REMON (TCP) and Infiniswap (RDMA).
Table~\ref{tab:remote_mem} summarizes their key differences.

\paragraph{\REMON} 
\REMON~\cite{remon_icde2026} transparently extends application memory over TCP/IP through a user-space allocator on the compute node and DRAM pools on memory nodes. It manages a reserved virtual-address region and fetches non-resident pages through a user-level fault handler over persistent connections. Under memory pressure, a background thread evicts LRU pages to the memory nodes by default.

We redirect DuckDB's allocation calls (e.g., \texttt{AllocateData} and \texttt{FreeData}) to \REMON's interfaces. The buffer manager treats \REMON pages as ordinary pointers and can use its explicit eviction and fetching APIs to enhance \emph{pin/unpin}. Thus, the operators' configurable block-level buffering translates into evict/fetch operations.

\paragraph{Infiniswap}
Infiniswap~\cite{Infiniswap2017} instead integrates remote memory with the kernel swap subsystem. Its block-device driver redirects OS swap I/O to remote memory over RDMA, while the kernel's page-replacement policy selects pages and Infiniswap moves them transparently.
Under memory pressure, the OS pages DuckDB buffers to the Infiniswap device, and later accesses trigger kernel-handled RDMA fetches.

\begin{table}[t]
  \centering
  \caption{Remote memory systems.}
  \label{tab:remote_mem}
  \small
  \begin{tabular}{lcc}
    \toprule
    & \textbf{\REMON} & \textbf{Infiniswap} \\
    \midrule
    Network transport  & TCP/IP       & RDMA (InfiniBand)  \\
    Implementation     & User space   & Kernel module      \\
    Integration point  & Allocator    & Swap device        \\
    Hardware required   & Commodity NIC & InfiniBand NIC    \\
  \bottomrule
  \end{tabular}
\end{table}


\section{Evaluation}
\label{sec:eval}

Similar to evaluations in prior work on memory-constrained query processing~\cite{Kuiper2025SavingPrivate, kuiper2024robust, Graefe1993Survey}, we evaluate \REMOP at both the operator and end-to-end workload levels. We first use controlled microbenchmarks to verify whether \REMOP's derived allocations reduce transfer rounds and runtime.
Then, we evaluate whether these gains carry into end-to-end TPC workloads and remain robust across various memory budgets, data sizes, network latencies, and swap backends.

\subsection{Experimental Setup}
\label{subsec:setup}

We use two CloudLab \texttt{c6220} nodes connected by TCP/IP and RDMA-enabled InfiniBand: one runs \REMOP-modified DuckDB and the other hosts the \REMON memory service.
Since \REMOP focuses on intra-operator buffer partitioning and swap batching, its benefits are largely orthogonal to the topology of memory nodes.
We therefore use a minimal setup with one compute node and one memory node.
Table~\ref{tab:setup} summarizes the hardware and network specs.

\begin{table}[t]
  \centering
  \caption{Experimental setup.}
  \label{tab:setup}
  \begin{tabular}{ll}
    \toprule
    Spec & Value \\
    \midrule
    Testbed & CloudLab \texttt{c6220}, Ubuntu 20.04 \\
    CPU & 16-core Intel Xeon E5-2650 v2 @ 2.60\,GHz \\
    DRAM & 64\,GiB physical; \REMON local cap: 1.05\,GB \\
    Network BW & 10\,GbE (TCP), 48.6\,GbE (RDMA) \\
    Network RTT & 0.155\,ms (TCP), 1.16\,$\mu$s (RDMA) \\
    RDMA & Mellanox ConnectX-3 InfiniBand \\
    \bottomrule
  \end{tabular}
\end{table}

\paragraph{Memory configuration}
Except for the memory sensitivity experiments, we keep DuckDB's buffer manager memory limit at 1\,GB and a per-operator memory budget of 64\,MB across 16 threads.
This cap places DuckDB in a memory-constrained regime where most experimental workloads reliably spill, making transfer-round overhead visible.
We set \REMON's local memory cap to 1.05\,GB for runtime headroom, provision 32\,GB of remote memory, and use a 256\,KB page size aligned with DuckDB's default block size.

\paragraph{Workloads}
For operator microbenchmarks, we use single-operator queries over synthetic data. Each tuple contains one \texttt{BIGINT} and one \texttt{VARCHAR} with a payload of about 1\,KB. To preserve memory pressure without a quadratic explosion in join output, we use the payload as the join key and set the \emph{join selectivity}, controlled by the key-domain size, as low as $1/256\text{K}$. For example, joining 256K and 1M tuples produces about 1M output tuples ($\sim$1\,GB).

For end-to-end evaluation, we use TPC-H (22 queries)~\cite{TPCH} and TPC-DS (99 queries)~\cite{TPCDS}, which represent analytical workloads with various operator types. Except in the scalability experiment, both run at SF10, which consistently triggers spilling under the 1\,GB budget while completing in reasonable time.
We mainly compare the \emph{spilling subsets}: 7 of 22 TPC-H queries and 13 of 99 TPC-DS queries.

\paragraph{Baselines}

We compare four configurations of the \REMOP DuckDB engine: \textbf{Vanilla} is unmodified DuckDB v1.0.0; \textbf{REMOP-heur} uses \REMOP's operators with disk-oriented heuristic knobs, isolating implementation overhead without tuned allocation; \textbf{REMOP(noPF)} uses the runtime-optimal policies without prefetching; and \textbf{REMOP} enables both policies and prefetching.
We also compare with \textbf{SPHJ}~\cite{Kuiper2025SavingPrivate}, a recent algorithmic EHJ optimization in DuckDB for graceful spilling, and use a $2{\times}2$ ablation to show that \REMOP's transfer-round optimization is complementary to this work. We further discuss it in \S\ref{subsec:disc} and \S\ref{sec:rel_work}.

\paragraph{Methodology}
We run each configuration five times with a $1{,}200$\,s timeout, and report the mean and standard deviation of the middle three runs excluding the fastest and slowest ones.
We profile runtime, swapped pages, and transfer rounds. Across multiple queries in TPC workloads, we use the geometric mean of their runtimes to reduce the influence of extreme values.

\subsection{Microbenchmarks}
\label{subsec:microbench}
 
\begin{figure}[t]
  \centering
  \begin{minipage}{0.49\linewidth}
    \centering
    \includegraphics[width=1.0\linewidth]{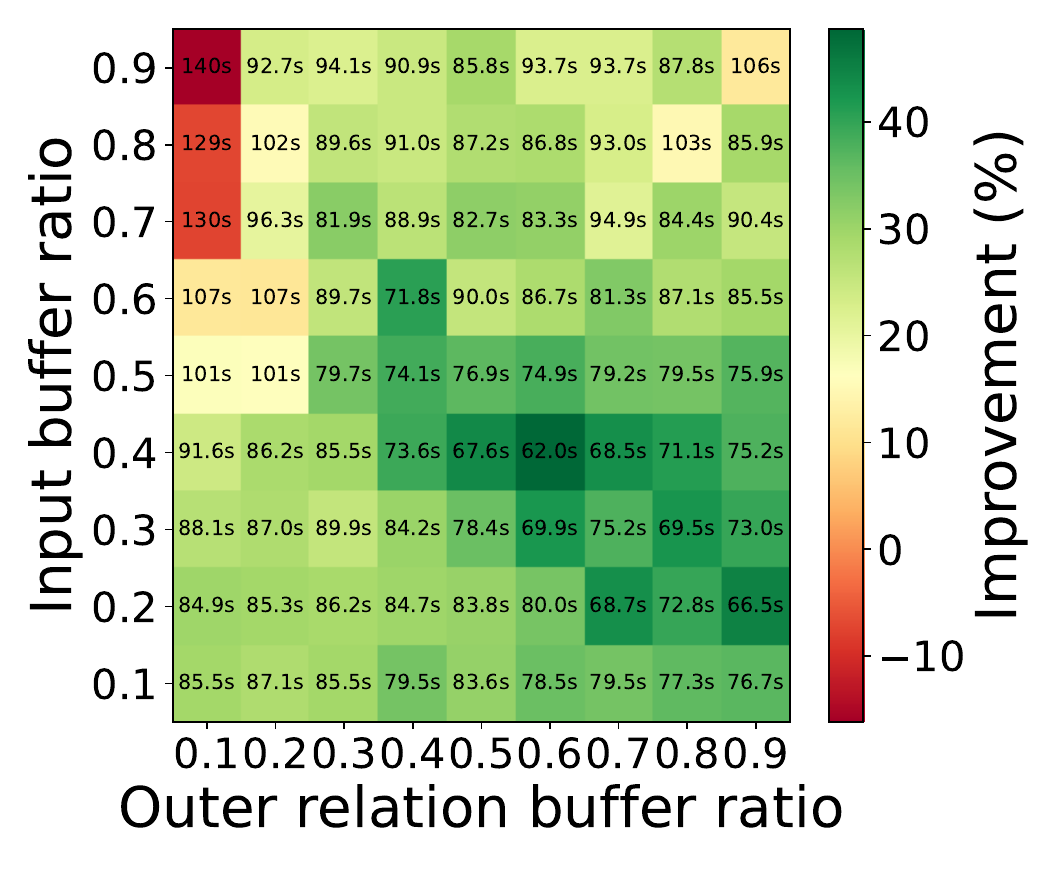}
  \end{minipage}\hfill
  \begin{minipage}{0.49\linewidth}
    \centering
    \includegraphics[width=1.0\linewidth]{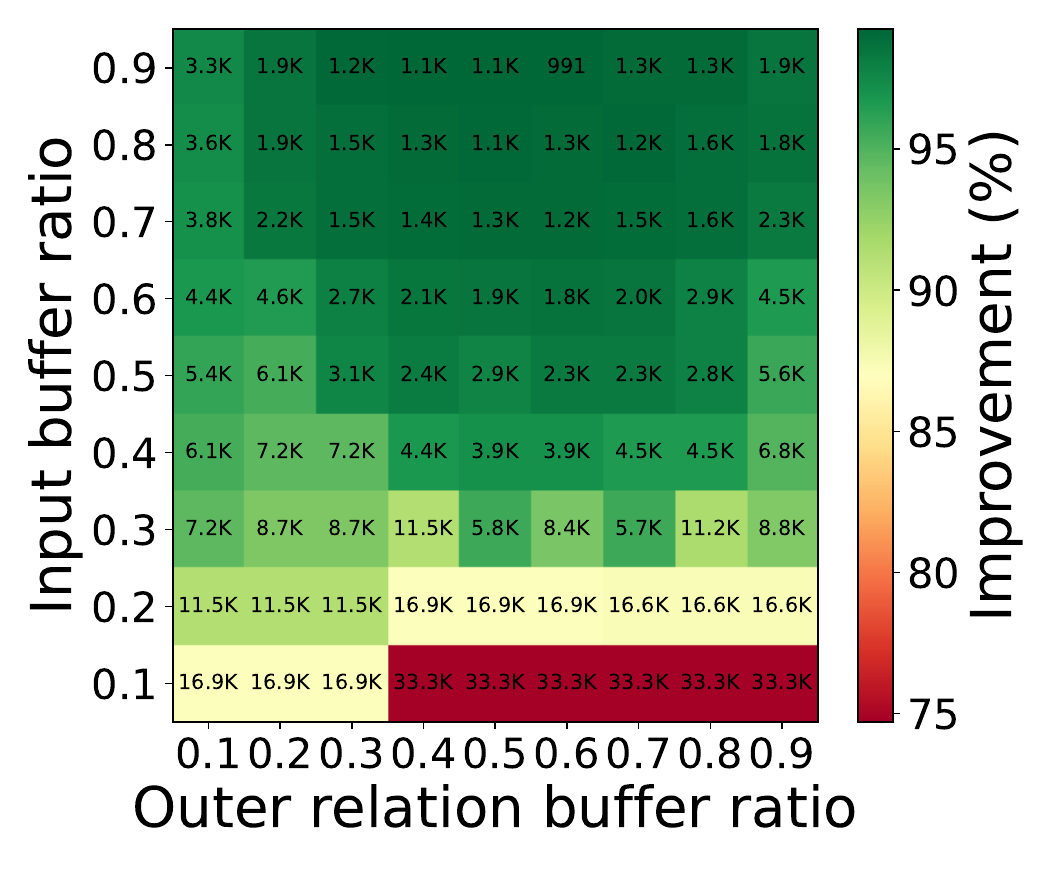}
  \end{minipage}
  \caption{\REMOP improvement (\%) over vanilla DuckDB for BNLJ. Left: runtime. Right: transfer rounds.}
  \Description{Two heatmaps showing \REMOP's improvement over vanilla DuckDB for blocked nested-loop join across buffer ratios: left reports runtime improvement and right reports transfer-round reduction.}
  \label{fig:nlj_heatmaps}
\end{figure}

\paragraph{Blocked Nested-Loop Join (BNLJ)}
\label{subsubsec:nlj_eval}

We first run a controlled synthetic nested-loop join to test whether \REMOP's BNLJ buffering strategy reduces transfer rounds and runtime.
For relations of 256K$\times$1M tuples, we vary the input-buffer ratio $r_{in}$ and the outer-relation buffer fraction $p_R$.
This sweep identifies an empirical runtime optimum and evaluates the analysis in \S\ref{subsec:nlj} before we apply the configuration to subsequent experiments.

Figure~\ref{fig:nlj_heatmaps} reports runtime improvement and transfer-round reduction across the buffer ratios.
Overall, allocations closer to a balanced split between the outer and inner relation buffers yield a much lower $C$ than strongly outer-heavy allocations.
With selectivity $f{=}1/256\text{K}$ and memory budget $M {=}4\text{K}$ pages, the minimum $C$ occurs at $r_{in}{=}0.9$, consistent with the analysis in \S\ref{subsec:nlj}.

Runtime is minimized at $(r_{in},p_R){=}(0.4,0.6)$.
Its $p_R$ lies between the $C$-minimizing split $p_R{=}0.5$ and the $D$-minimizing limit $p_R{\to}1$, reflecting the trade-off captured by $L$ for the measured network parameter $\tau$.
At this $L$-optimal point, \REMOP improves runtime by 48.7\% and reduces transfer rounds by 97.1\%.
We therefore use $(r_{in},p_R){=}(0.4,0.6)$ for BNLJ in the remaining experiments.

\begin{figure}[t]
  \centering
  \begin{minipage}{0.435\linewidth}
    \centering
    \includegraphics[width=1\linewidth]{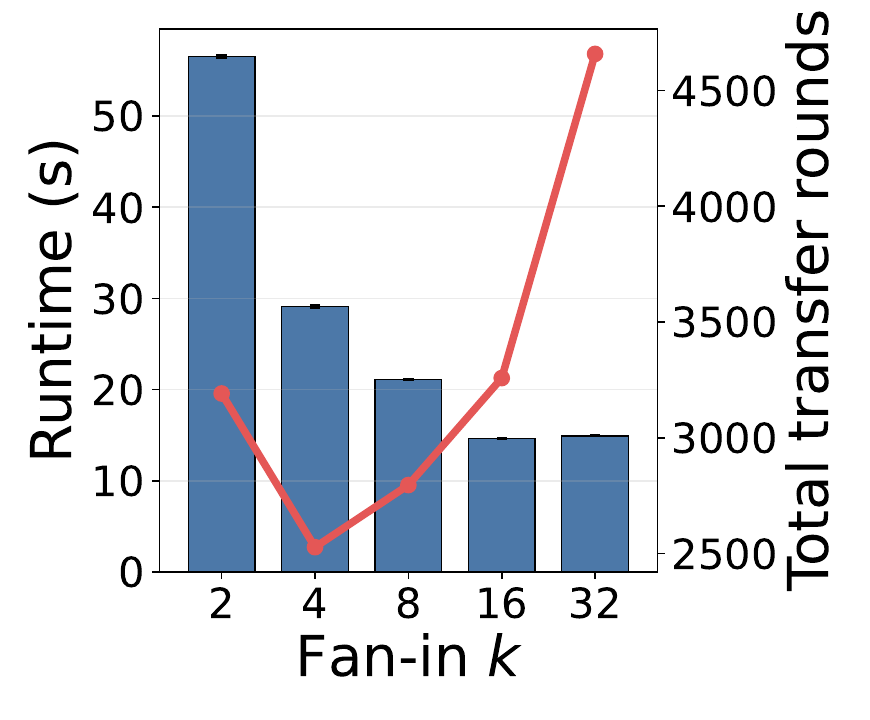}
  \end{minipage}\hfill
  \begin{minipage}{0.555\linewidth}
    \centering
    \includegraphics[width=1\linewidth]{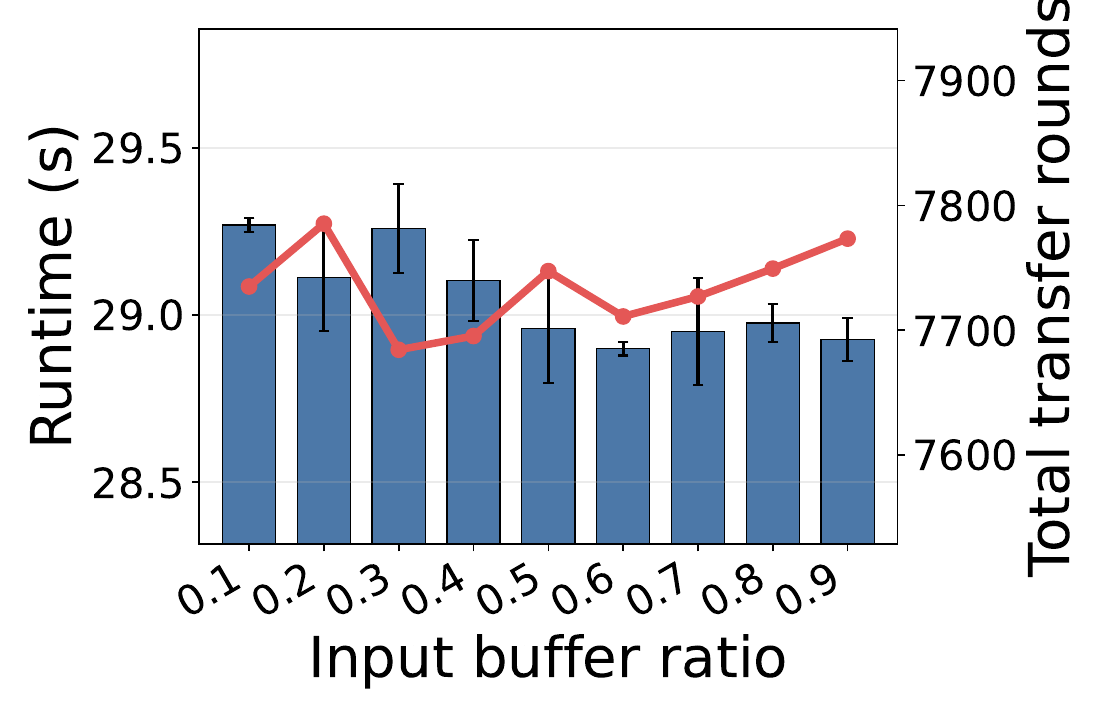}
  \end{minipage}
  \caption{EMS performance over fan-in $k$ and input-buffer ratio.}
  \Description{Two plots comparing vanilla DuckDB and \REMOP for an external sort microbenchmark, varying merge fan-in $k$ (left) and input-buffer ratio (right).}
  \label{fig:sort_microbench}
\end{figure}

\paragraph{\texorpdfstring{$k$}{k}-Way External Merge Sort (EMS)}
\label{subsubsec:kway_eval}
To evaluate \REMOP's EMS optimization, we run a synthetic \texttt{ORDER BY} query over 1M tuples under the same memory budget and compare vanilla DuckDB's two-way merge sort with our $k$-way merge implementation.

Figure~\ref{fig:sort_microbench} (left) varies the fan-in $k$ with input-buffer ratio fixed at 0.6.
Transfer rounds are minimized at $k{=}4$, consistent with our analysis.
Runtime instead reaches its minimum at $k{=}16$, because increasing fan-in reduces the number of merge passes and the transfer volume $D$, initially outweighing the moderate increase in transfer rounds $C$.
The improvement tapers for $k{\ge}16$, as the reduction in merge passes saturates and coordination overhead becomes more visible.
This behavior is captured by $L$, whose runtime optimum shifts from the $D$-minimizing limit $k{=}M-1$ toward the $C$-minimizing $k{=}4$.

Figure~\ref{fig:sort_microbench} (right) fixes $k{=}4$ and varies the input-buffer ratio.
Runtime is minimized near $r_{in}{=}0.6$, while both runtime and transfer rounds vary little across the ratios, indicating less sensitivity to this knob than to fan-in $k$. 
Combining the two parameter sweeps, we use the runtime-optimal configuration $(k,r_{in}){=}(16,0.6)$ for EMS in the remaining experiments.

\paragraph{External Hash Join (EHJ)}
\begin{figure*}[t]
  \centering
  \begin{subfigure}{0.55\linewidth}
    \centering
    \includegraphics[width=\linewidth]{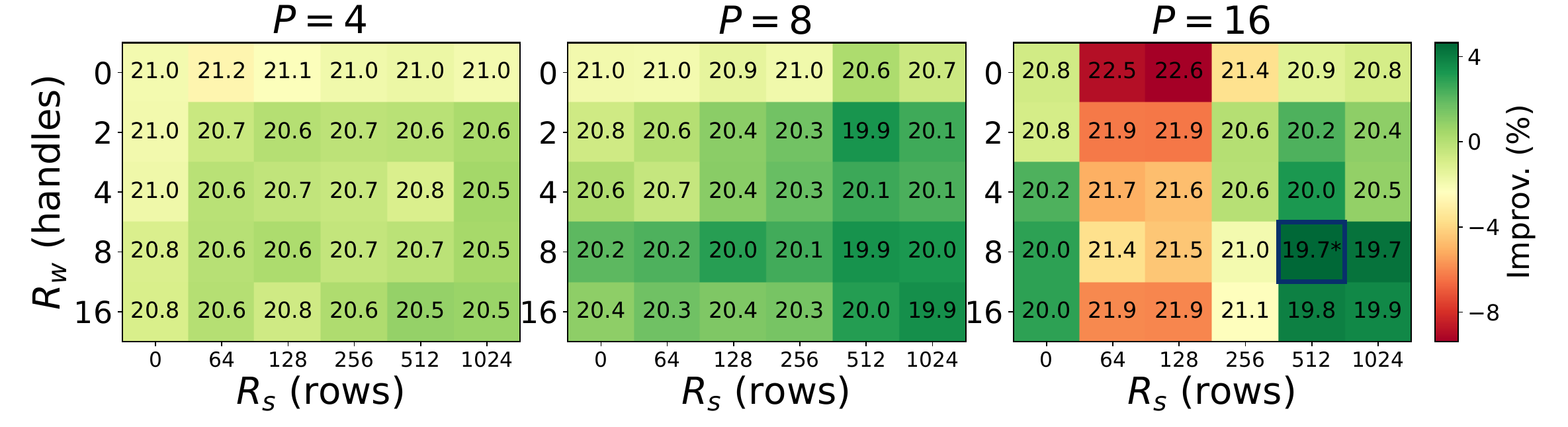}
    \caption{EHJ runtime improvement (\%) across build-write handles $R_w$, probe staging rows $R_s$, and partition count $P$.}
    \label{fig:hj_heatmaps}
  \end{subfigure}\hfill
  \begin{subfigure}{0.21\linewidth}
    \centering
    \includegraphics[width=\linewidth]{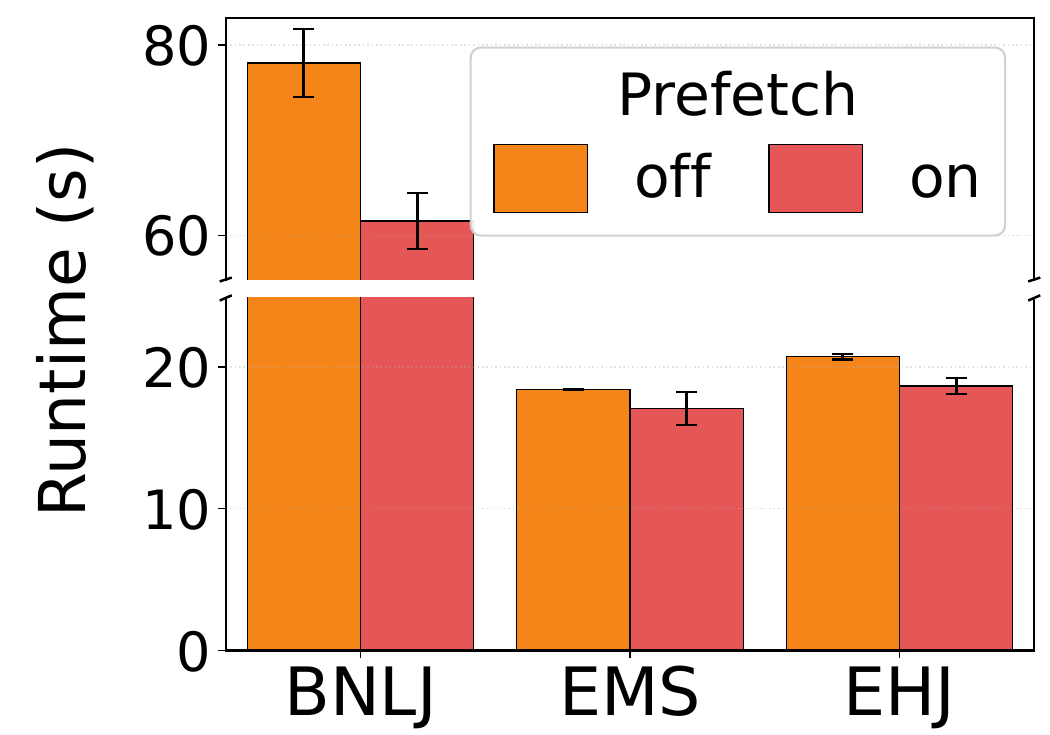}
    \caption{Per-operator prefetch buffer ablation.}
    \label{fig:prefetch}
  \end{subfigure}\hfill
  \begin{subfigure}{0.21\linewidth}
    \centering
    \includegraphics[width=\linewidth]{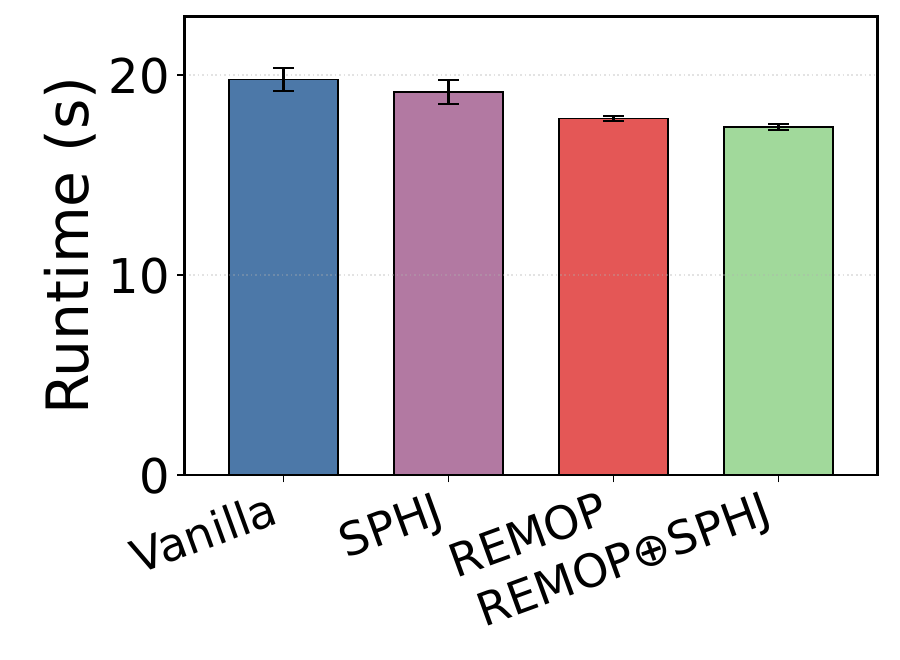}
    \caption{SPHJ vs.\ \REMOP ablation on EHJ.}
    \label{fig:sphj_ablation}
  \end{subfigure}
  \caption{Operator-level microbenchmarks.}
  \Description{Three panels: EHJ knob-sweep heatmaps, the per-operator prefetch ablation bars, and the SPHJ-versus-REMOP ablation bars.}
  \label{fig:micro}
\end{figure*}

To evaluate \REMOP's EHJ optimization, we run a single hash join under the fixed 1\,GB budget.
Unlike BNLJ, EHJ avoids a quadratic number of comparisons, so we scale the relations to 16M$\times$64M tuples (64\,B per row) to induce spilling.

We vary the build-write pool $R_w$ (implemented as pinned page handles per build partition) and the probe staging pool $R_s$ (measured in staged rows) for different partition counts $P\in\{4,8,16\}$. 
Figure~\ref{fig:hj_heatmaps} reports the runtime improvement over vanilla DuckDB.
A value of zero denotes DuckDB's default, with no additional pinned build-partition handles or staged probe rows. 
Across all the partition counts, increasing either $R_w$ or $R_s$ from zero generally lowers runtime by batching build-partition writes and spilled-probe flushes, respectively.
As $P$ grows, lower-runtime configurations favor more write-side capacity relative to the read buffer, consistent with the $1{:}\sigma\sqrt{P}$ read-to-write and read-to-staging ratios derived in \S\ref{subsec:hj}.
The overall optimum, $P{=}16$ with $R_w{=}8$ and $R_s{=}512$ (boxed), reduces runtime by 4.6\%.
We use this configuration in the remaining experiments.

\paragraph{Prefetch Buffering}
Beyond buffer partitioning, \REMOP overlaps remote transfers with computation through a per-operator prefetch double buffer.
With each operator using its selected configuration, we compare runtime with prefetching disabled and enabled.
As Figure~\ref{fig:prefetch} shows, prefetching reduces runtime by 21.3\% for BNLJ, 10.0\% for EHJ, and 7.4\% for EMS.
The benefit is largest for BNLJ because its repeated scans of the inner relation provide a predictable stream with substantial prefetching opportunities, whereas it is smallest for EMS as its sequential merge stream already hides more of the transfer latency.

\paragraph{SPHJ Ablation}
To test whether \REMOP complements the prior algorithmic hash-join optimization SPHJ, we port SPHJ to our \REMOP-modified DuckDB and run a$2{\times}2$ ablation on the synthetic hash join.
As Figure~\ref{fig:sphj_ablation} shows, SPHJ alone reduces runtime by 3.1\%, while \REMOP's buffer-partitioning and prefetch policies alone reduce it by 9.8\%.
Enabling both yields an 11.9\% reduction, exceeding either optimization alone and confirming that they are complementary.
In the remaining experiments, the \REMOP configuration excludes SPHJ, while the SPHJ baseline excludes \REMOP.

\subsection{End-to-End Benchmark}
\label{subsec:e2e_eval}
To evaluate whether \REMOP's operator-level improvements translate to end-to-end analytical workloads, we run TPC-H and TPC-DS at scale factor 10 (SF10) under the same 1\,GB memory budget. We use the optimal configurations identified in the microbenchmarks (BNLJ: $r_{in}=0.4$, $p_R=0.6$; EMS: $k=16$, $r_{in}=0.6$; EHJ: $P=16$, $R_w=8$, $R_s=512$). This tests both the generality of the optimizations and the robustness of these settings across different query plans.
To better isolate the advantage of \REMOP, we focus on the TPC queries that spill to the remote memory.

\begin{figure*}[t]
  \centering
  \includegraphics[width=\textwidth]{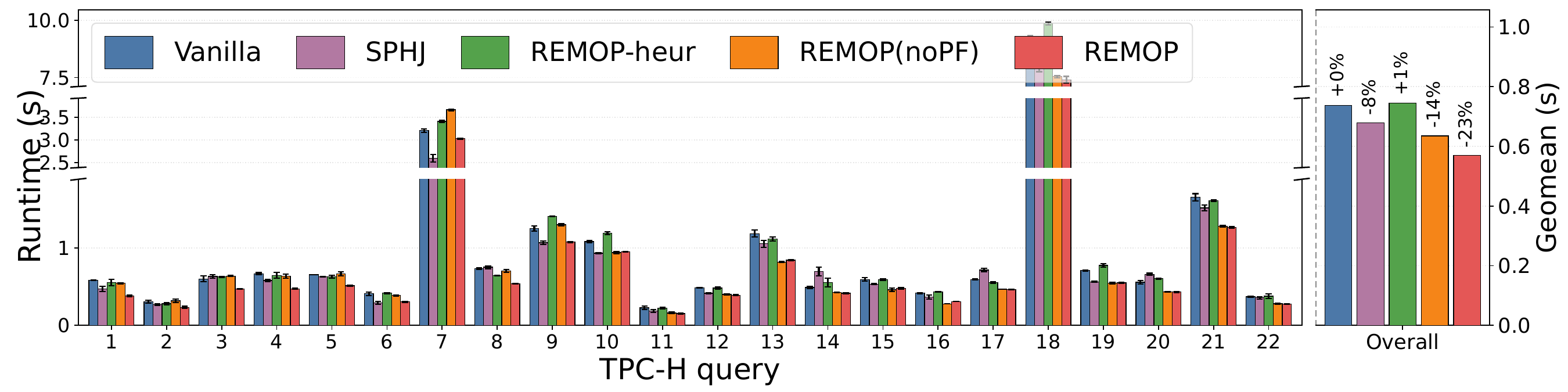}
  \caption{TPC-H SF10 per-query runtime with the overall geometric mean.}
  \Description{Bar chart of TPC-H SF10 per-query runtimes for vanilla DuckDB and \REMOP, with a line indicating percentage improvement.}
  \label{fig:tpch_sf10}
\end{figure*}

\paragraph{TPC-H}
Figure~\ref{fig:tpch_sf10} reports per-query runtime and improvement on the TPC-H benchmark.
Across all 22 queries, \REMOP reduces the geometric-mean runtime by 22.7\% relative to vanilla, versus 13.9\% without prefetching. It also outperforms the SPHJ baseline, while the classical disk-oriented heuristic configuration shows little overhead over vanilla DuckDB.
In particular, on the spill-heavy queries Q7, Q9, Q10, Q13, Q18, Q19, and Q21, \REMOP reduces the geometric mean by 20.2\%.

\begin{figure*}[t]
  \centering
  \includegraphics[width=\textwidth]{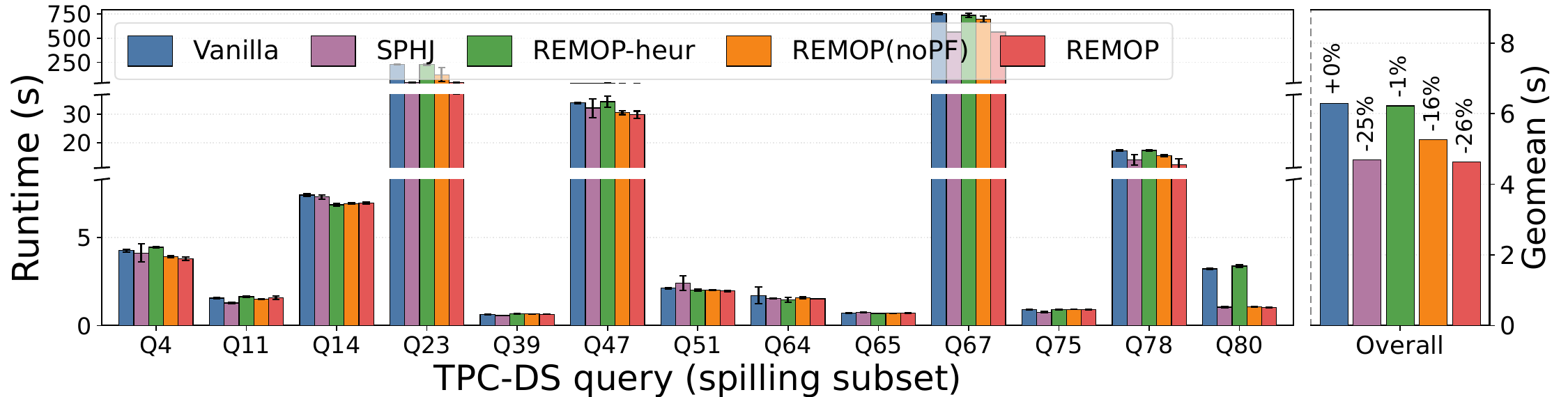}
  \caption{TPC-DS SF10 runtime on spilling queries with the overall geometric mean.}
  \Description{Grouped-bar chart of TPC-DS SF10 per-query runtime for the five configurations on the spilling subset, with a geometric-mean panel.}
  \label{fig:tpcds_sf10}
\end{figure*}

\paragraph{TPC-DS}
Figure~\ref{fig:tpcds_sf10} reports per-query runtime of the spilling queries on the TPC-DS benchmark.
Over the spilling subset (Q4, Q11, Q14, Q23, Q39, Q47, Q51, Q64, Q65, Q67, Q75, Q78, and Q80), \REMOP reduces the geometric-mean runtime by 26.4\% relative to vanilla DuckDB, which is comparable to the SPHJ baseline (25.4\%) and ahead of the prefetch-free variant (16.3\%). This confirms that \REMOP's operator-level mechanisms remain effective across more varied analytical plans.

\subsection{Memory Sensitivity}
\label{subsec:mem_eval}
\begin{figure*}[ht]
  \centering
  \includegraphics[width=\textwidth]{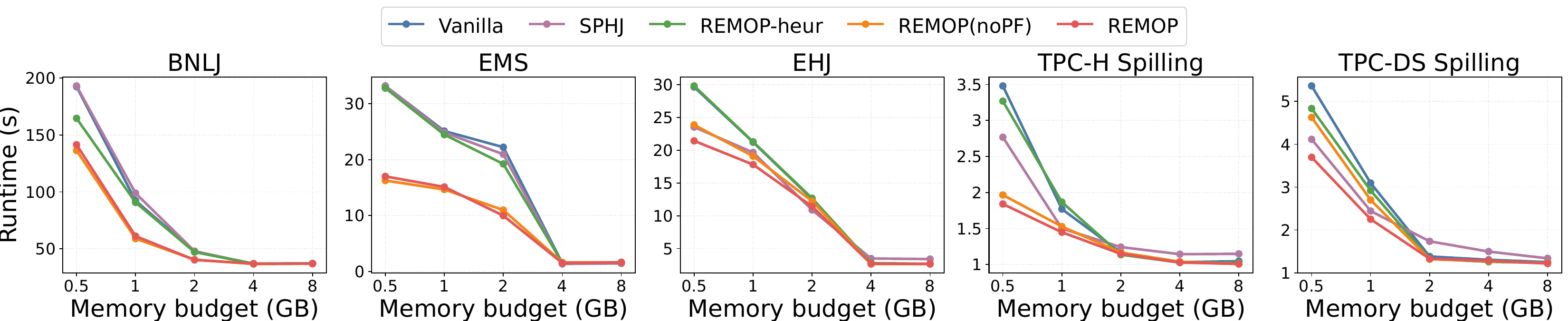}
  \caption{Runtime sensitivity to the local memory budget $M$.}
  \Description{Five line charts of runtime versus memory budget for the five configurations on BNLJ, EMS, EHJ, and the TPC-H/TPC-DS spilling subsets.}
  \label{fig:mem_sensitivity}
\end{figure*}
\REMOP targets memory-constrained execution, so its advantage is expected to grow with spilling pressure and diminish when the working set fits locally. To verify this, we hold input sizes fixed and vary the local memory budget $M$ over $\{0.5,1,2,4,8\}$\,GB.
The \REMON local cap is set to $(M{+}0.05)$\,GB accordingly.

Figure~\ref{fig:mem_sensitivity} confirms this trend. At $M=512$\,MB, \REMOP reduces runtime by 26\% (BNLJ), 49\% (EMS), 28\% (EHJ), 47\% (TPC-H spilling), and 31\% (TPC-DS spilling) relative to vanilla DuckDB. The configurations converge as $M$ grows and spilling subsides, showing that the gains arise from reducing remote-memory transfer cost.

\subsection{Scalability}
\label{subsec:scale}
To evaluate how \REMOP's benefit scales with data size, we hold the memory budget fixed at 1\,GB and increase the input size.
This raises the $N/M$ ratio, transfer volume, and number of transfer rounds, allowing us to test whether \REMOP's transfer-round savings persist or compound as the working set outgrows local memory.

\begin{figure*}[t]
  \centering
  \includegraphics[width=\textwidth]{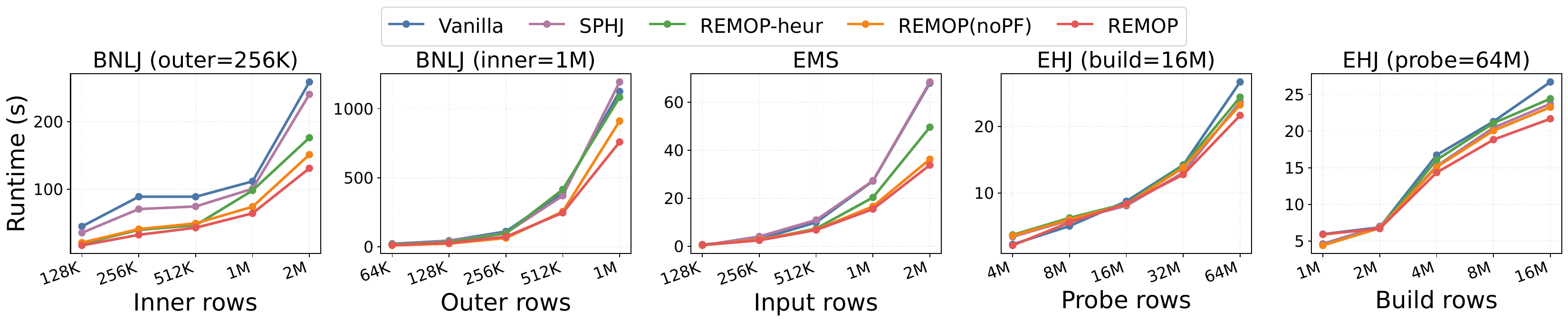}
  \caption{Scalability per operator.}
  \Description{Five line charts of runtime versus relation or input size for the five configurations on BNLJ (two panels), EMS (one panel), and EHJ (two panels).}
  \label{fig:synth_scalability}
\end{figure*}

For each operator, we vary one input dimension at a time while holding the others fixed.
As Figure~\ref{fig:synth_scalability} shows, \REMOP's advantage persists and generally widens with input size.
It reduces BNLJ runtime by 33\%\textendash{}64\% and EMS runtime by up to 50\% at 2M tuples, while EHJ sees a more modest reduction of up to 19\%.
The growing benefit across the three operators shows that \REMOP's transfer-round savings become more prominent as spilling pressure rises.

\begin{figure}[t]
  \centering
  \includegraphics[width=0.9\linewidth]{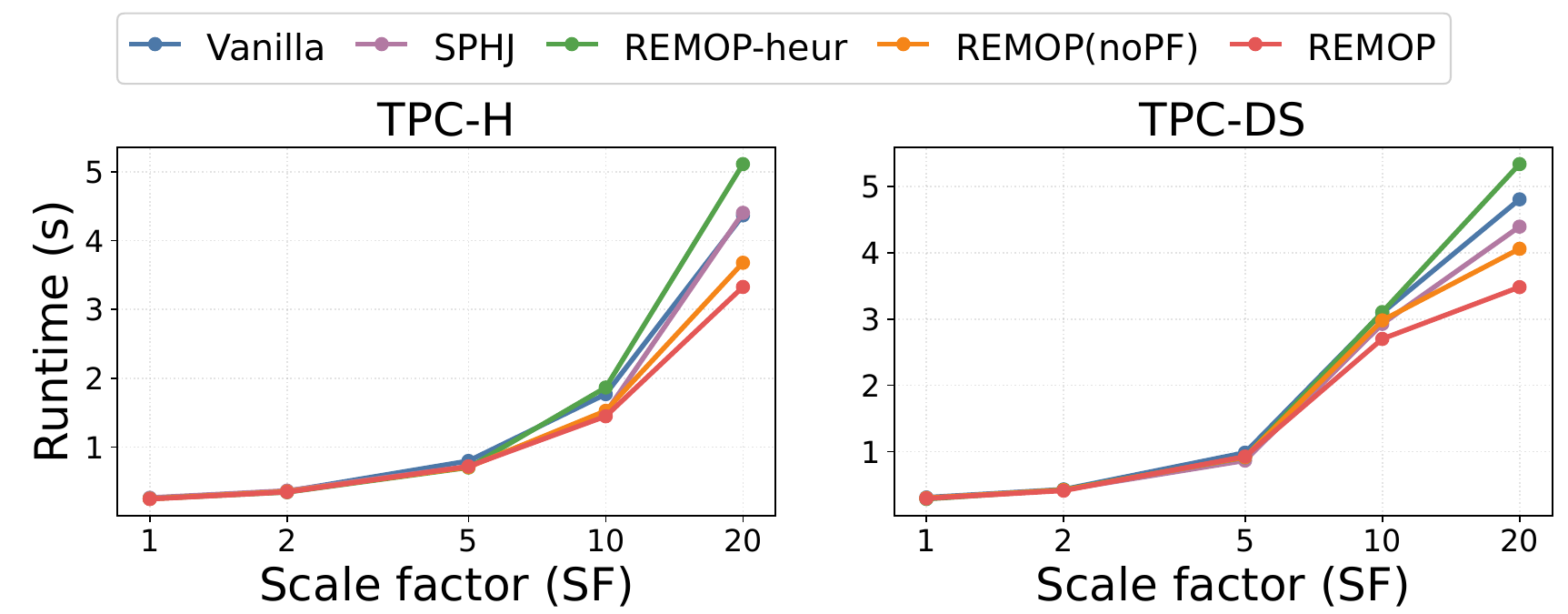}
  \caption{End-to-end TPC scalability.}
  \Description{Two line charts of geometric-mean runtime versus scale factor for the five configurations on the TPC-H and TPC-DS spilling subsets.}
  \label{fig:tpc_scalability}
\end{figure}

We next evaluate scalability on the end-to-end benchmarks.
As Figure~\ref{fig:tpc_scalability} shows, the TPC workloads follow the operator-level trend: \REMOP's lead over vanilla DuckDB widens with scale factor, reaching 24\% on TPC-H and 28\% on TPC-DS at SF20.
Thus, the per-operator transfer-round savings translate into a growing end-to-end advantage as the datasets grow.

\subsection{Network Latency Sensitivity}
\label{subsec:network_eval}

\begin{figure*}[t]
  \centering
  \includegraphics[width=0.95\textwidth]{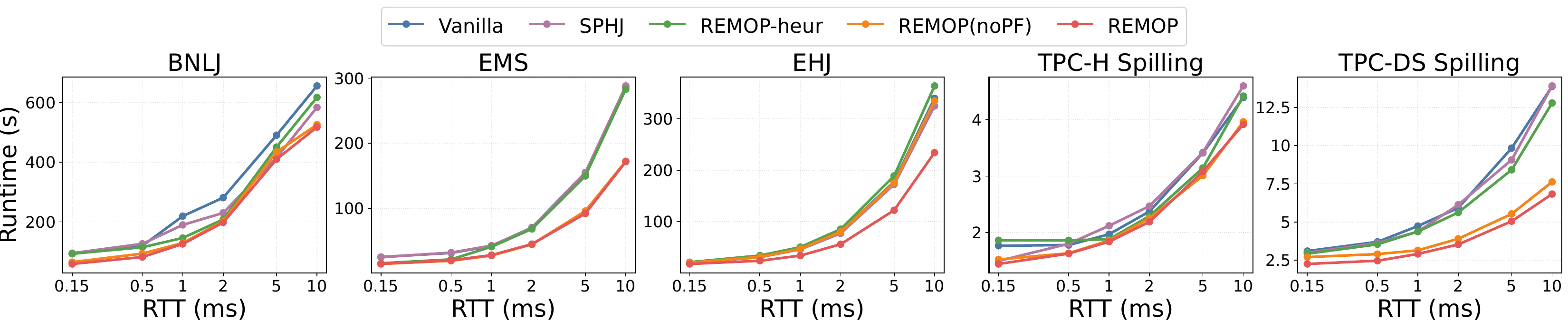}
  \caption{Sensitivity to network RTT.}
  \Description{Five line charts of runtime versus network RTT for the five configurations on BNLJ, EMS, EHJ, and the TPC-H/TPC-DS spilling subsets.}
  \label{fig:lat_sensitivity}
\end{figure*}

To test how \REMOP behaves as remote-memory access becomes more expensive, we evaluate its sensitivity to increasing network latency. 
This models practical deployments with higher RTTs (e.g., cross-datacenter remote memory) and allows us to quantify how well \REMOP mitigates them.
With an underlying network latency of 0.15\,ms, we use Linux \texttt{tc netem}~\cite{TcNetemManpage} to introduce additional delay on both the compute and memory nodes, yielding RTTs from 0.15\,ms to 10\,ms with 20\% uniform jitter.

As Figure~\ref{fig:lat_sensitivity} shows, end-to-end runtime rises with RTT, and \REMOP's advantage widens as remote access becomes more expensive. Specifically, at an RTT of $10$\,ms, \REMOP reduces runtime by $21\%$ (BNLJ), $40\%$ (EMS), $31\%$ (EHJ), $11\%$ (TPC-H spilling), and $51\%$ (TPC-DS spilling) relative to vanilla DuckDB.
As RTT increases, the larger network parameter $\tau$ makes the transfer-round term a greater component of latency cost $L$. The widening advantage therefore supports our cost model and shows that \REMOP mitigates latency growth by reducing transfer rounds.

\subsection{Swap-Backend Comparison}
\label{subsec:swap_eval}

\begin{figure*}[t]
  \centering
  \includegraphics[width=0.95\textwidth]{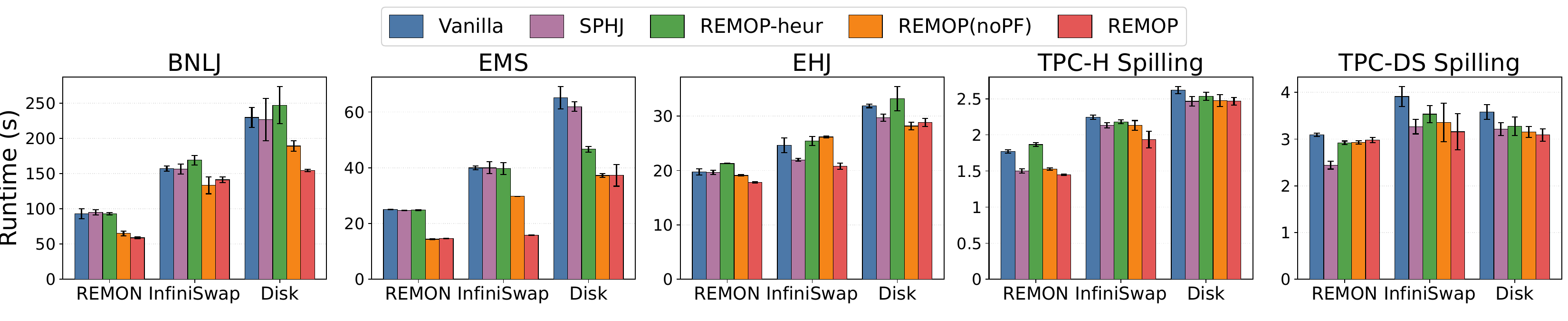}
  \caption{Swap-backend comparison.}
  \Description{Five grouped-bar charts of runtime for the five configurations across the REMON, InfiniSwap, and Disk backends on BNLJ, EMS, EHJ, and the TPC-H/TPC-DS spilling subsets.}
  \label{fig:swap_comparison}
\end{figure*}

To further demonstrate the generality of \REMOP, we evaluate it under three different swap backends: \REMON (TCP-based user-space remote memory), Infiniswap (RDMA-based kernel swap), and local disk spill (DuckDB temporary files).
All runs use the same query settings aligned with the prior experiments.

As shown in Figure~\ref{fig:swap_comparison}, \REMOP consistently improves runtime across all three backends with different bandwidth and latency characteristics.
Moreover, under our memory cap, using \REMON as the swap backend outperforms disk-based spilling across all workloads, highlighting the effectiveness of user-space remote memory and the advantage of \REMOP when RTT is non-negligible.
In particular, although raw RDMA latency/bandwidth is substantially better than TCP network and disk (Table~\ref{tab:tier_numbers}), the Infiniswap backend does not uniformly outperform \REMON and disk spilling. We discuss the plausible underlying causes in \S\ref{subsec:disc}.
 
\subsection{Discussions}
\label{subsec:disc}

\noindent\textbf{SPHJ orthogonality:} SPHJ~\cite{Kuiper2025SavingPrivate}, building on the authors' prior work on robust out-of-memory operators~\cite{kuiper2024robust}, changes EHJ's algorithmic behavior so build overflow spills through a unified, compressible buffer pool instead of triggering a slower fallback. \REMOP leaves the join algorithm unchanged and instead optimizes an orthogonal dimension: it sizes the buffers in each phase to minimize transfer rounds and adopts prefetching to further reduce the round-trip cost.
The two techniques therefore compose rather than overlap. Figure~\ref{fig:sphj_ablation} shows that their combined 11.9\% improvement exceeds either alone, demonstrating that \REMOP can be layered on algorithmic hash-join optimizations such as SPHJ.

\noindent\textbf{RDMA integration overheads:}
Although RDMA offers significantly lower latency and higher bandwidth (as shown in Table~\ref{tab:setup}), these numbers are measured at the network level and do not directly translate to application-level performance. 
In practice, applications need additional software wrapping to leverage RDMA, such as Infiniswap, which provides a kernel-level swap interface over RDMA. 
Triggering kernel swap introduces overheads such as page-fault handling and swap-in/out bookkeeping, and the database buffer manager does not directly control kernel swapping decisions.
Additionally, Infiniswap operates at the OS page granularity (typically 4\,KiB), which mismatches DuckDB's 256\,KiB block size, potentially increasing the number of swap events and amplifying per-round overheads.
Together, these effects can increase end-to-end latency and diminish the raw RDMA advantage.
In contrast, \REMON provides a user-space remote-memory substrate with a lighter integration path and less software overhead.
Results in \S\ref{subsec:swap_eval} show that although \REMOP still outperforms vanilla DuckDB with Infiniswap, this backend can trail \REMON and even local disk spilling.

\section{Related Work}
\label{sec:rel_work}

\paragraph{Disaggregated and Remote Memory}
Memory disaggregation decouples compute from network-accessible memory, spanning early shared-memory architectures~\cite{Lim2009Disaggregated}, LegoOS's split-kernel resource management~\cite{Shan2018LegoOS}, OS-level RDMA swapping in Infiniswap~\cite{Infiniswap2017}, application-level far-memory abstractions in AIFM~\cite{Ruan2020AIFM}, and CXL pooling for cloud platforms and DBMSs~\cite{Pond2023,Ahn2024CXLHANA}.
Unlike systems requiring specialized RDMA or CXL hardware, our prior work \REMON~\cite{remon_icde2026} provides remote memory through TCP/IP on commodity cloud instances without kernel modifications. We use it as the primary backend and Infiniswap to evaluate \REMOP across substrates.

\paragraph{Memory-Aware Query Processing}
Prior work studies memory allocation across queries and operators.
Graefe surveys multi-pass sort and hash execution under insufficient memory~\cite{Graefe1993Survey}.
Mehta and DeWitt redistribute memory dynamically among concurrent memory-intensive queries~\cite{MehtaDeWitt1993}, while Nag and DeWitt allocate memory across operators in complex decision-support plans~\cite{NagDeWitt1998}.
Recent work from Otaki et al.~\cite{Otaki2025PagedMemoryManagement} provides explicit paged memory management for adaptive query execution, and Laser~\cite{HuangLi2024Laser} learns buffer-aware scheduling under master--standby replication.
These approaches primarily optimize \emph{inter-operator} allocation and scheduling; \REMOP instead optimizes \emph{intra-operator} buffers for remote-memory transfer granularity and latency.

\paragraph{Database Buffer Management}
Database buffer managers determine page residency and access.
LeanStore~\cite{Leis2018LeanStore} reduces lookup overhead through pointer swizzling and optimistic latching. Umbra~\cite{Neumann2020Umbra} uses virtual memory for transparent disk spilling. Anti-caching~\cite{DeBrabant2013AntiCaching} gives the DBMS explicit control over evicting cold tuples.
Whereas these systems determine \emph{which} pages remain local and \emph{how} they are accessed, \REMOP, at a different level, controls remote-memory-aware allocation and access patterns within individual operators.

\paragraph{Operator-Level Optimizations}
Join processing under varying memory sizes has long been studied~\cite{Shapiro1986Joins}.
For hash joins, Zeller and Gray adapt to changing availability in multi-user environments~\cite{ZellerGray1990}, Davison and Graefe respond to memory contention~\cite{DavisonGraefe1994}, and Jahangiri et al.~\cite{Jahangiri2022HybridHash} examine robust dynamic hybrid designs.
Kuiper et al.~\cite{Kuiper2025SavingPrivate} more recently reduce degradation when build inputs exceed RAM by coordinating memory across concurrent hash joins.
Besides the join operator, external-sort research includes classical multiway-merge analysis~\cite{Knuth1998TAOCP3}, external-memory I/O bounds~\cite{AggarwalVitter1988IOComplexity}, practical techniques such as replacement selection, read-ahead, and multi-phase merging~\cite{Graefe2006Sorting}, and adaptation to changing memory availability~\cite{PangCareyLivny1993MemoryAdaptiveSorting,ZhangLarson1997DynamicMemoryMergesort}.
Balkesen et al.~\cite{BalkesenSortHash2013} show that hardware and internal parameters can determine the sort--hash trade-off.
\REMOP studies the analogous effect of transfer latency and granularity in remote memory.

Specifically, \REMOP differs primarily in its cost model.
Classical operator analyses emphasize total I/O volume or operations, which suits disk and bandwidth-dominated settings. In remote memory, transferring the same bytes in more rounds can incur substantially more latency.
\REMOP therefore models transfer rounds and derives operator-specific allocations that may trade data volume for fewer round trips under a fixed local budget.
This focus complements paged query execution~\cite{Otaki2025PagedMemoryManagement}, learned buffer-aware scheduling~\cite{HuangLi2024Laser}, out-of-memory GPU streaming~\cite{Wang2025GraphDataAccess}, and broader operator-level cost modeling~\cite{Wu2020OperatorLevelCostModeling}.


\section{Conclusions}
\label{sec:conclu}

We presented \REMOP, a remote-memory-aware operator optimization framework that reduces out-of-memory query latency.
\REMOP formalizes a latency cost model that incorporates transfer rounds and provides a general intra-operator allocation mechanism.
We demonstrated it on blocked nested-loop join, external merge sort, and external hash join as representative case studies and implemented it in DuckDB over a network-based remote-memory backend.

Across single-operator microbenchmarks and end-to-end TPC benchmarks, \REMOP reduces transfer rounds by up to $97\%$ and lowers the runtime of spilling TPC-H and TPC-DS queries by $22.7\%$ and $26.4\%$ over vanilla DuckDB, with the advantage widening as remote-memory latency grows. These results show that modeling transfer rounds, rather than data volume alone, is key to efficient out-of-memory query processing over remote memory.

\section*{Acknowledgment of AI-Generated Content}
In this work, the authors used AI-assisted tools including OpenAI ChatGPT and Anthropic Claude to facilitate prototype system development, experiment running, and paper writing. The authors reviewed, edited, and verified all such output. All final contents including analyses, experimental data, and conclusions were produced and checked by the authors.

\balance
\bibliographystyle{IEEEtran}
\bibliography{citation}

\end{document}